\documentclass[smallextended]{svjour3}       
\smartqed  
\usepackage{graphicx}
\usepackage{cite}
\usepackage{longtable} 
\usepackage[hyphens]{url}
\usepackage[pdftex,breaklinks]{hyperref}
\hypersetup{
colorlinks=true,
citecolor=blue,
linkcolor=blue,
filecolor=blue,
urlcolor=blue
}            

\long\def\symbolfootnote[#1]#2{\begingroup%
\def\thefootnote{$\fnsymbol{footnote}$}\footnotetext[#1]{#2}\endgroup} 
\begin{document}

\markboth{D. Helbing and P. Mukerji}{Crowd Disasters as Systemic Failures}

%
%

\title{Crowd Disasters as Systemic Failures: Analysis of the Love Parade Disaster\protect\footnotemark[1]}

\author{Dirk Helbing$^{1,2,3}$ \and Pratik Mukerji$^1$}

\institute{$^1$ Chair of Sociology, in particular of Modeling and Simulation, ETH Zurich, Swiss Federal Institute of Technology, Clausiusstrasse 50, 8092 Zurich, Switzerland\\
$^2$ Risk Center, ETH Zurich, Swiss Federal Institute of Technology, 
Scheuchzerstrasse 7, 8092 Zurich, Switzerland\\
$^3$ Santa Fe Institute, 1399 Hyde Park Road,  Santa Fe, NM 87501, USA}
\symbolfootnote[1]{The following complementary webpage with time-ordered, geo-located videos been set up for this paper: \protect\url{http://loveparadevideos.heroku.com/}} 

\date{Received: date / Accepted: date}

\maketitle

\begin{abstract}
Each year, crowd disasters happen in different areas of the world. How and why do such disasters happen?
Are the fatalities caused by relentless behavior of people or a psychological state of panic that makes the crowd `go mad'? Or are they a tragic consequence of a breakdown of coordination? These and other questions are addressed, based on a qualitative analysis of publicly available videos and materials, which document the planning and organization of the Love Parade in Duisburg, Germany, and the crowd disaster on July 24, 2010. Our analysis reveals a number of misunderstandings that have widely spread. We also provide a new perspective on concepts such as `intentional pushing', `mass panic', `stampede', and `crowd crushs'. The focus of our analysis is on the contributing causal factors and their mutual interdependencies, not on legal issues or the judgment of personal or institutional responsibilities. Video recordings show that people stumbled and piled up due to a `domino effect', resulting from a phenomenon called `crowd turbulence' or `crowd quake'. This was the consequence of amplifying feedback and cascading effects, which are typical for systemic instabilities. Hence, things can go terribly wrong in spite of no bad intentions from anyone. Comparing the incident in Duisburg with others, we give recommendations to help prevent future crowd disasters. In particular, we introduce a new scale to assess the criticality of conditions in the crowd. This may allow preventative measures to be taken earlier on. Furthermore, we discuss the merits and limitations of citizen science for public investigation, considering that today, almost every event is recorded and reflected in the World Wide Web.
\end{abstract}

\section{Introduction}

Crowd disasters are known since at least the Roman Empire. As a consequence, building codes for stadia were developed. The Coliseum in Rome, Italy, which is considered to be one of the seven world wonders, is probably the best known example of Roman building experience.
While it could take up between 50,000 and 73,000 visitors, it had 76 numbered entrances, and visitors exited through the same gate through which they had entered. In fact, exits were located side by side, around the entire circumference of the Coliseum. As a consequence, the Coliseum could be evacuated within just 5 minutes, an efficiency that is not even reached by modern stadia due to their smaller number of exits. 
\par
Building codes and regulations for mass events have also been written and updated after recent crowd disasters, such as the ones in Bradford (1985) or Hillsborough, Sheffield (1989) \cite{control1,control2,control3,control4,control5,control6,control7,control8}. Today's knowledge about the dynamics of crowds is considerable and summarized in Refs. \cite{Predtechenskii,Fruin,Still,evacuation,TranSci,schadenc,encyklopedia}. Furthermore, a lot of experience in organizing safer mass events has recently been gained from the organization of religious pilgrimage \cite{crowdsafety1,crowdsafety2,crowdsafety3,crowdsafety4,crowdsafety5}. In recent years, there is also a quickly growing body of literature on evacuation experiments \cite{exp1,exp2,exp3,exp4,exp5,exp6,exp7,exp8,schadenc,encyklopedia} and pedestrian simulations \cite{sim1,sim2,sim3,sim4,sim5,sim6,sim7,sim8,sim9,sim10,sim11,sim12,evacuation,exp6}, and various related commercial software products are now available. Thus, how was it possible that 21 people died and more than 500 were injured during the Love Parade on July 24, 2010? 
\par
A crucial point for the safety of mass events is that they are (or at least should be) organized in a way that is robust against many kinds of disturbances (such as weather conditions, human errors, etc.). This is why the organization of a mass event includes the elaboration of contingency plans. Why then can crowd disasters still happen? 
\par
This paper will reveal that the Love Parade disaster was not the result of a single mistake. We will rather show that the Love Parade disaster resulted from the interaction of {\it several} contributing factors. It is probably the first time that a detailed analysis can be performed with publicly available documents: not just investigation reports by public authorities \cite{Polizeibericht,FinalReport} and the media, but also maps from Google Earth \cite{Earth} and 360 degree photographs \cite{360}, videos accessible through YouTube \cite{youtube}, documents released by Wikipedia \cite{Wiki,Wikipedia,wikikarte} and Wikileaks \cite{Leaks}, and other sources. In some sense, this opens up a new age of public investigation. However, to avoid misunderstandings,  we would like to underline that our analysis focuses on the course of events and causal interdependencies among them, while they do not draw any conclusions regarding legal issues or personal or institutional responsibilities, which must be judged by other experts (see, for example, Ref. \cite{responsible}). 
\par
The remainder of this paper is structured as follows: Section 2 provides an overview of the situation before and during the Love Parade disaster. This includes a historical background, a description of the festival area (including in- and outflows), and a timeline reconstructed from many video recordings. Section 3 will analyze various factors contributing to the disaster, while Section 4 will focus on causal interdependencies and interaction effects. Section 5 discusses our findings and Section 6 concludes with lessons learned for the organization of future mass events. The novelty of this paper is four-fold: it concerns (1) the structured analysis of large amounts of publicly available video recordings of a disaster, (2) the interpretation of the disaster as a systemic failure (where the interaction of various factors created a systemic instability, causing an overall loss of control), (3) a revision of common views about crowd disasters, and (4) the introduction of a scale reflecting the criticality of crowd conditions (and proposed counter-measures). 

\section{Overview of the Situation}

The following section will try to give a short overview of the situation during the Love Parade in Duisburg and the planning beforehand. A large number of documents are
now publicly available (see Ref. \cite{Dok} for a collection of links). This includes the planning documents \cite{Leaks}, the event log of the regulatory authority of the city of Duisburg \cite{Einsatz}, and the evacuation analysis \cite{Entfluchtung}. Publicly accessible materials and eye witness reports now amount to several hundred pages \cite{wordpress} and more than 500 video recordings \cite{Videos}. This useful collection of materials is the result of the efforts of many volunteers. It is certainly not possible (but also not the purpose) of this article to give a complete representation of materials. We will rather focus on the most relevant details in order to avoid a distraction of the reader from the main factors that have contributed to the disaster. 
\par
The interested reader is invited to gain a more complete picture himself or herself, based on the media reports provided in Refs. \cite{Spiegel,Tagesschau,WDRmediathek} and documentaries of several TV channels \cite{SpiegelTV,SpiegelVideoDoku,RTL}. The view of the organizer is presented in Ref. \cite{LopaventVideos}. Further video documentations are available from private persons \cite{overv,syncro,VideoDoku,Critique}. An interpretation of the events, overlayed to a satellite picture, can be found in Ref. \cite{Satellite}. 
\par
In order to make an independent assessment possible, our own analysis will largely refer to authentic materials that are publicly accessible. 
Videos of a subset of surveillance cameras are available until 16:40 \cite{Surveillance}. Timelines can be found in Refs. \cite{timeline,Polizeibericht,Einsatz}. 
Complementary to this article, we provide a time-ordered and geo-located collection of videos from visitors of the Love Parade \cite{videocoll}. A YouTube channel with videos of the Love Parade exists as well \cite{youtube}. The collection \cite{Videos} contains further videos. Many of these videos have been synchronized \cite{Zeitstrahl,pizzamanne}, and some of them have been cut together in the form of multi-view videos documenting the course of events \cite{Pizzamanne}. A set of highly relevant private videos around the time of the disaster can be found in Refs.  \cite{kaydee271,backtony,real02,hitower,Todesparade,goonies,coolwojtek,mbreezer,pizzamanne,hell,rkjorge70}. 
\par
Note that, when referring to secondary sources (such as public media reports), we will sometimes use wordings such as ``apparently'' or ``seems to'', in order to indicate that access to primary sources would be desirable for an in-depth analysis. 

\subsection{History of the Love Parade}

The Love Parade is a popular electronic dance music festival in Germany that was first organized in Berlin in 1989, and annually repeated in the same city until 2003. The events in 2004 and 2005 had to be cancelled because of funding problems and a coordinated opposition of political parties (e.g. related to the waste resulting from the event) \cite{Wiki}. In 2006, the parade made a comeback with the support of a fitness studio. The Love Parade in summer 2007 was again planned for Berlin, but the event was cancelled, since the Senate of Berlin did not issue the necessary permits on time. After negotiations with several German cities, it was then decided to move the Love Parade to the Ruhr Area, an agglomerate of major German cities, in the next years. The first of these events took place in Essen on August 25, 2007, with 1.2 million visitors. In July 2008, it was organized in Dortmund. The 2009 event, planned for Bochum, was cancelled due to security concerns, particularly as a critical situation had apparently occurred the year before \cite{Wikipedia}. The last Love Parade took place on July 24, 2010, in Duisburg, where 21 people died and more than 500 were injured in a crowd disaster. The chain of events underlying this disaster will be analyzed in the following sections.

\subsection{Description of the Festival Area} \label{festa}

The festival area of the Love Parade in 2010 was approximately 100,000 square meters large \cite{Spiegel} 
and located in the area of a previous freight station of the city of Duisburg. For a 360 degree view of the festival area and its surroundings see Ref. \cite{360}. In contrast to the open area concept of the Love Parade in Berlin (see the picture in Ref. \cite{Berlin}), the annual Carnival in Cologne, and the 20th World Youth Day gathering with the Catholic Pope in 2005 in Cologne-Marienfeld, Germany \cite{Pope}, the festival area was constrained by railway tracks on the East and by a freeway on the West. In response to concerns from the regulatory authority that the area would be too small for the expected number of up to 1.4 million expected visitors  \cite{Spiegel}, the city of Duisburg combined its late approval of the event with the condition to restrict the number of concurrent visitors to 250,000.
\par
To overcome security issues seen by the regulatory authority (there was some discussion to cancel the event overall), the organizer of the Love Parade decided to fence the whole festival area. This moved the responsibility to the building regulatory agency \cite{Spiegel} and required the event to satisfy the ``Versammlungsst\"attenverordnung'' \cite{control8}, which is the German safety regulation for the organization of mass events. However, there were still concerns that the standard safety requirements would not be met. It is conceivable that these concerns were not fully considered due to a desire to approve the event \cite{political}, particularly as Duisburg was nominated as Germany's `cultural capital' of the year, and the opinion prevailed that the Love Parade would make the cultural program and the city more attractive \cite{Spiegel}. To overcome the concerns, an expert opinion was requested from a prominent crowd researcher. The report argued that the festival area could be sufficiently well evacuated in an emergency situation \cite{Entfluchtung}. However, the study did not analyze normal entry and exit conditions in detail.
\par
\begin{figure}[htbp]
\begin{center}
\includegraphics[width=12cm]{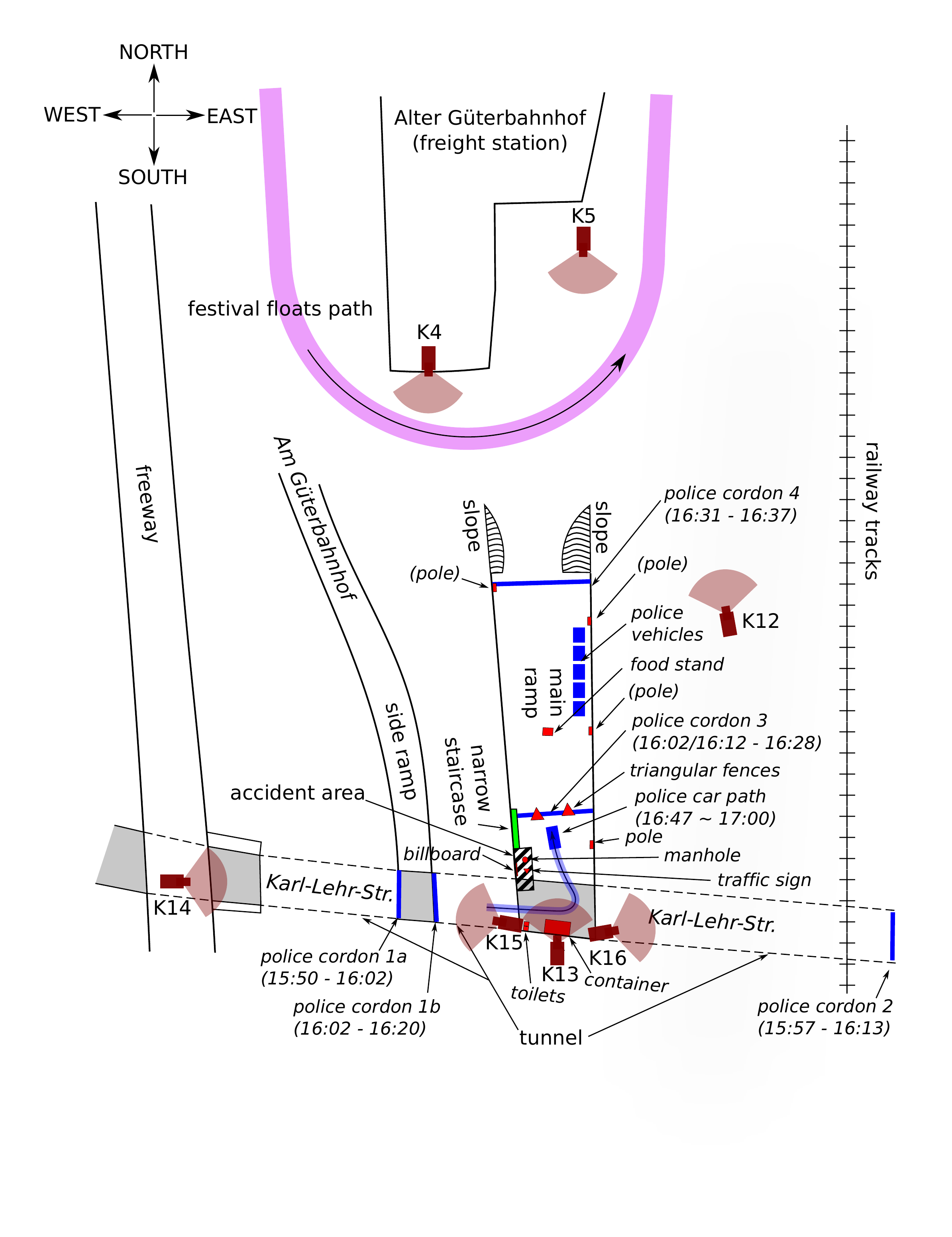}
\end{center}
\caption{Illustration of the festival area and the ways to and from the area. Camera positions are shown as well as locations and events that are relevant for the analysis of this study. Note that the indicated timing of the police cordons (as reconstructed from video recordings) slightly differs from the police report \cite{Polizeibericht}, but the differences are small.}
\label{fArea}
\end{figure}
Figure \ref{fArea} gives an overview of the festival area. It shows that the festival area could be entered only via a tunnel, ``Karl-Lehr-Stra{\ss}e'', which also served as the only exit from the area. In the middle of that tunnel, there is the main ramp that leads to the festival area. The tunnel and the ramp together determine an inverse T-shaped geometry of in- and outflows. A side ramp in the West (``Am G\"uterbahnhof'') was assigned as an additional exit ramp \cite{Polizeibericht}, but basically not used. The smallest overall diameter of the tunnels in the East and in the West was about 20 meters \cite{wikikarte}. The ramp itself was 26 meters wide and 130 meters long \cite{Spiegel}. 
Based on the maximum flow value of 1.225 persons per meter per second \cite{Weidmann}, this would imply a hypothetical maximum flow of 114,660 persons per hour and a density of 1.75 persons per square meter, if the entire ramp width was usable. However, the actual capacity was significantly lower than this due to the following factors (see also Sec. \ref{KStill}):
\begin{enumerate}
\item The maximum possible flow is inconvenient and potentially unsafe, and therefore not suited as a basis for planning \cite{Predtechenskii,HCM,LOS1,LOS2}. 
\item Counterflows are expected to reduce the capacity by $6-14\%$ \cite{Weidmann}, resulting in a maximum hypothetical flow of 98,608 persons per hour. 
\item The 90 degree turn to and from the tunnels is expected to reduce the capacity as well.
\item Walking in groups reduces the capacity further \cite{GROUPS}.
\item Alcohol and drugs are expected to have a negative impact on capacity as well.
\item A considerable amount of capacity must have been lost due to fences \cite{Fences}, a food stand \cite{foodstand}, and vehicles on the side of the ramp \cite{policecars}.
\end{enumerate}
The flow model of the organizer assumed the following numbers \cite{flowmodel}:
\par\begin{table}[htbp]
\begin{center}
\begin{tabular}{|l|c|c|}
\hline
Time & Expected inflow/h & Expected outflow/h \\
\hline
14:00-15:00 & 55,000 & 10,000 \\ 
15:00-16:00 & 55,000 & 50,000 \\
16:00-17:00 & 55,000 & 45,000 \\
17:00-18:00 & 90,000 & 55,000 \\
\hline
\end{tabular}
\end{center}
\caption{Expected inflows and outflows estimated by the organizers (see Ref. \cite{flowmodel} for more details). Based on these values, the maximum number of visitors on the festival area was expected to be 235,000 (while a capacity of 250,000 was approved and more than 1 million visitors were expected, according to announcements before and during the event \cite{Schaller}). Estimates based on surveillance videos of camera 13 suggest that the actual flows were considerably below the values in the above table. According to Ref. \cite{Bewegungsmodell}, the inflow  in the time period
between 14:00 and 15:00 varied between 280 and 600 persons per minute and the outflow between 6 and 80 persons per minute. 
Between 15:00 and 15:40, it  varied between 450 and 750 persons per minute and the outflow between 40 and 250 persons per minute. 
This is 30-50\% below expectations of the organizer of the Love Parade and implies a maximum number of visitors on the festival area of about 175,000.}
\label{flows}
\end{table}
According to Table \ref{flows}, between 17:00 and 18:00 the organizers expected an inflow of 90,000 and an outflow of 55,000 people, which could not have been handled by the wide ramp without the use of suitable crowd control. Problems had to be expected already for much smaller flow rates, as there were vehicles and a food stand as well as fences on the ramp, which must have reduced its capacity considerably. This risk factor certainly had to be carefully considered by the crowd management concept. In fact, the side ramp (see Fig. \ref{fArea}) was attributed as an additional exit ramp, and the organizational concept foresaw the possibility to reduce the visitor flows through `isolating devices' (access control points), which were located in front of the tunnel entrances \cite{AccesS}. Despite this, access control was given up intermittently because of the large pressure from incoming visitors (see Ref. \cite{Polizeibericht} and Tables \ref{FIRST} to \ref{LAST}). The festival area itself was apparently not overcrowded (see caption of Table \ref{flows} and aerial photographs \cite{aerial,Hoist}).
So, why and how did the crowd disaster happen in the inverse T-section formed by the tunnel and the ramp, even though the visitor flows were apparently smaller than expected (see Table \ref{flows}) and a more than 3,000 people strong police force was on duty? To address this question, we will first present an expert opinion on  the crowd disaster. Then, we will summarize the course of events, and analyze the contributing factors in more detail. 

\subsection{Expert Report by Prof. Dr. G. Keith Still} \label{KStill}

An expert report dated December 9, 2011, which became public in February 2012 \cite{StillReport}, analyzes the implications of the flow model presented in Table \ref{flows}. In the following, we summarize the essence of this report in our own words: 
\begin{enumerate}
\item Safe crowd conditions can be usually assumed for densities up to 2-3 persons per meter and minute and a maximum acceptable flow of 82 persons per meter and minute (which is considerably below the maximum possible flow) \cite{Fruin2}.
\item All areas, in which higher crowd densities may occur or where many people may accumulate, must be analyzed for risks.
\item The safety concept must list those risks and also, who is responsible to handle them. The organizational structure (in particular, who takes what kinds of decisions) must be fixed before the event. Particular attention must be paid to crowd management and communication (loud speakers, signs, maps and other plans).
\item All authorities involved in the organization of the event are responsible for the safety of the crowd. The division of responsibility should be regulated in the concept of the event. A mass event should not be approved, if it does not satisfy the applicable safety regulations.
\item The reason for most crowd disasters in the past was a failure to regulate the flow of people in high throughput areas. 
\item The organization of an event needs plans for normal operation, but also contingency plans for all kinds of incidents.
\item There are basically three ways of influencing the safety of crowds: design, information, and crowd management.
\item At the Love Parade in Duisburg, the capacity of the main ramp to and from the festival area was given by the minimum usable width of the ramp. Due to two triangular fence structures \cite{Fences,TRIANGLE}, which were apparently not shown in the maps,  the effective width of the ramp was only 10.59 meters. According to the expert report, this implies a maximum safe flow of 10.59  
meters $\times$ 82 persons per meter and minute $\times$ 60 minutes = 52,103 persons per hour. However, the maximum expected flow between 17:00 and 18:00 was 145,000 persons per hour, which would require a width of 29.5 meters. Therefore, at the Love Parade in Duisburg, problems with the in- and outflows and a critical accumulation of people had to be expected. 
\item Once the crowd density exceeds between 4 or 5 persons per square meter, congestion can build up quickly, which implies high risks for people to stumble or fall (particularly if the ground is uneven). Therefore, injuries can easily happen. 
\item People in a dense crowd cannot see what happens a few meters away from them, and they are not aware of the pressure in front.
\item The density, noise, and chaos in a dense crowd cause a natural desire to leave the crowd. Due to a lack of suitable crowd control and guidance, visitors of the Love Parade in Duisburg could only see a narrow staircase as a possible emergency exit (see Fig. \ref{fArea}). When trying to get there, the pressure towards the staircase increased and eventually triggered the crowd disaster.
\end{enumerate} 
The analysis of the effective capacity of the main ramp suggests that problems on the ramp were foreseeable, and the question arises, why the obstacles were placed there. However, a complete assessment should also consider the existence of the side ramp (see Fig. \ref{fArea}). Moreover, 
due to the applied access control, the flows on the main ramp did not reach the expected flows by far.  This can be directly concluded from the fact that there was never any significant congestion between the two triangular obstacles defining the narrowest part of the ramp, before the flow was controlled in this area from 16:02 on; this is clearly visible in the surveillance videos \cite{TRIANGLE}. An active bottleneck, in contrast, would be characterized by the formation of a queue \cite{RMP}. 
\par
Queues of people did not form in the middle of the ramp, but rather at the upper end, where visitors were trying to enter the festival area. This, however, is not the location where the crowd disaster happened. Therefore, while one {\it had} to expect problems in the middle of the ramp where the triangular obstacles were located, the crowd disaster was actually {\it not} caused by those obstacles. The course of events that resulted in the crowd disaster involved many contributing factors, as we will show in the following. This conclusion of our study is in line with a quote referring to the Hillsborough disaster of 1989, which apparently goes back to the Archbishop of York and can be found in Keith Still's PhD thesis \cite{Still}:  ``Events of the magnitude of Hillsborough don't usually happen just for one single reason, nor is it usually possible to pin the blame on one single scapegoat... Disasters happen because a whole series of mistakes, misjudgements and mischance happen to come together in a deadly combinations.'' This should be kept in mind when Keith Still's expert report on the crowd disaster in Duisburg points out that it is merely based on the evidence presented to him and that it answers only the questions posed to him.

\subsection{Timeline}

The chronology presented in Table \ref{FIRST} is an abbreviated version of the timeline that was originally provided by the organizers of the Love Parade together with their documentary movie \cite{LopaventVideos}. It is largely supported by the surveillance videos \cite{Surveillance} and other public sources. 
Additional points will be discussed afterwards.
\par{\footnotesize\begin{table}[htbp]
\begin{tabular}{p{1.5cm}p{9.7cm}}
\hline
12:02 & The festival area is opened. Visitors can enter the area via the access control points from East and West via the tunnel. \\
13:00 & The inflow is reduced by closing 10 of 16 isolating devices, both on the East and West entrance towards the tunnel. \\
13:45-14:15 & No important disturbances or queues of visitor flows occur in the entry area.\\
Around 14:00 & Official start of the Love Parade.\\
14:15-14:30 & The concentration of visitors increases at the end of the entrance ramp towards the festival area (due to obstructions by `floats', i.e. moving music trucks).\\ 
14:30-15:15 & The crowd manager tries to order support by the police. The organizer states that the person responsible for connecting to the police (the `liaison officer') did not have a working walkie talkie or mobile phone.\\
14:30-15:06 & The visitor flow on the ramp and from the West increases.\\
Around 15:00 & Reduction of the visitor flow by closing as many isolating devices as possible.\\
15:12-15:34 & Change of police shifts \cite{PoliceShifts}. 5 police cars drive into the ramp area.\\
15:31 & Visitors ignore the fence on the side of the main ramp, following police forces, who have temporarily opened it. Shortly later, visitors overcome fences also on the other side of the ramp, which should prevent them from taking the steep slope up to the festival area.\\
15:50 & A first chain of police forces (police cordon) is formed in front of the side ramp, blocking in- and outflows in the West \cite{CordonWest} (see cordon 1a in Fig. \ref{fArea}). \\
15:50-15:57 & A second police cordon closes the tunnel to the East (see cordon 2 in Fig. \ref{fArea}).\\
Around 16:02 & There is a sudden strong visitor flow towards the festival area from the West. The first police cordon is moved behind the side ramp (see cordon 1b in Fig. \ref{fArea}).\\ 
From 16:02 & Police forces start to control the flows to (and from) the festival area in the middle of the ramp (where the ramp is narrowest due to some fences) \cite{Cordon3}. Queues start to form on both sides of the resulting bottleneck \cite{QUEUES}. \\ 
Around 16:06 & There are just a few visitors between the three police cordons.\\
Around 16:07 & A jam of visitors forms in the West part of the tunnel. \\
Around 16:09 & A jam of visitors forms above the chain of police forces on the ramp, when trying to exit the festival area.\\
16:12-16:28 & The third police cordon is completed (see cordon 3 in Fig. \ref{fArea}). It stops the in- and outflows completely, where the fences narrow down the ramp. \\
Around 16:13 & The small ramp is opened as entrance to the festival area. Visitors climb over fences. \\
Around 16:14 & The second police cordon in the East opens up, and visitors enter the area of the big ramp from below \cite{secondcordon}.  \\
Around 16:17 & First visitors try to enter the festival area via a narrow staircase connecting the lower part of the ramp with the festival area on top \cite{firststaircase}. Afterwards, the staircase is blocked by two security people \cite{block}.\\ 
Around 16:21 & The first police cordon in the West dissolves \cite{wordpress,firstdissolves}. The previously waiting visitors move towards the ramp and encounter there the dense flow of visitors coming from the East.\\
16:22 & First people climb the pole \cite{wordpress,firstpole}.\\
16:22-16:24 & The third police cordon still keeps the ramp closed, while the pressure increases from both sides (i.e. inflow and outflow).\\
16:24-16:28 & The third police cordon is dissolved \cite{thirddissolves}. \\ 
Around 16:27 & The narrow staircase is used by people to get up to the festival area \cite{stairuse}. Someone climbs on top of a traffic sign \cite{signclimb}.\\
16:31-16:37 & A fourth police cordon is formed in the upper area of the ramp \cite{forthform}. At the same time, the density in the lower area of the ramp increases steadily.\\
After 16:40 & The situation gets out of control. More and more visitors try to get up to the festival area via the small staircase, the pole and a container (used by the crowd management, located at the lower end of the ramp in the South). \\
\hline
\end{tabular}
\caption{Timeline according to the organizers of the Love Parade.}
\label{FIRST}
\end{table}}
The video recordings of the surveillance cameras and the related chronology, which were publicly provided by the organizer,  end at 16:40 
(``in respect of the victims''). Tables \ref{SECOND} and \ref{LAST} present additional information that is relevant for a reconstruction of the causes of the crowd disaster. A time-ordered, geo-coded video link collection supplementing this paper allows the readers to gain an independent impression \cite{videocoll}. 
\par{\footnotesize\begin{table}[htbp]
\begin{tabular}{p{1.5cm}p{9.7cm}}
\hline
8:03 & The police receive an e-mail informing them about the official approval of the Love Parade  \cite{wordpress}. \\ 
Until 12:00 & The construction work (leveling work) of a bulldozer on the festival ground takes longer than planned and delays the opening of the festival for approximately one hour \cite{Spiegel}. \\
13:33 & 20,000 techno fans are waiting in the West and are creating a lot of pressure to get in \cite{Spiegel,mauer}.\\ 
13:44 & The police are worried that the access point may be overrun \cite{Spiegel}.\\ 
Around 14:00 & A police officer asks the crowd management to make a loudspeaker announcement, but this cannot be done, because there is no working loudspeaker equipment despite requirements to have one \cite{Spiegel}.\\
After 14:03 & Visitors are obstructed by floats (music trucks), while trying to enter the festival area from the ramp  \cite{Spiegel,rampflow}. \\
14:42 & The obstruction by the floats on the festival area causes a jam of arriving visitors on the ramp almost up to the tunnel \cite{Spiegel}.\\
14:52 & For some time, it is not possible to enter the festival area from the ramp \cite{wordpress}.\\ 
15:06 & The minister of interior visits the crisis management team \cite{wordpress}.\\ 
15:30-18:00 & Mobile phones do not work due to an overload of the mobile phone networks \cite{Spiegel}.\\
From 15:31 & Visitors start to climb the slope in the West of the main ramp and one minute later in the East to get to the festival area \cite{Boeschung}. \\
Around 16:00 & Turmoil and critical crowd conditions occur in front of the access points. A policeman instructs the crowd management to open the access point in the West \cite{Spiegel}. The access point in the East is intermittently opened to reduce the pressure in the crowd \cite{Spiegel}.\\
16:31 & A fence at the West side of the tunnel is opened to allow an emergency vehicle to enter. Hundreds of visitors make use of the occurring gap to enter the tunnel \cite{Polizeibericht}.\\
Around 16:30 & Visitors overcome fences in the tunnel \cite{overcomingfences}. \\
16:35-16:43 & People scream for help and shout at others they should hurry up; some seem to panic, but others try to calm them down; the situation changes quickly: people change between screaming and laughing; some people manage to climb the staircase, but there is still no continuous flow of people on the staircase \cite{interrupted}. People scream they are about to die \cite{scream}. The traffic sign is already bent \cite{signbent}. People shout from above that those on the narrow staircase should move on \cite{moveon}.\\ 
Around 16:36 & Crowd turbulence and critical situation around the pole \cite{lightpole}. \\
Starting 16:38 & Police are limiting the number of people on the staircase (usually 2 or 3 at a time), but make sure that people do not stop on the staircase 
\cite{limiting}.\\
Around 16:40 & An unconscious women is passed on to the narrow staircase and elevated up \cite{unconscious}. A sparse, slowly moving crowd in the tunnel moves towards the festival area \cite{tunnelcrowd}.  \\ 
Starting approx. 16:40 & Police cars in the city make loudspeaker announcements that the festival area is completely full and will not be accessible to further visitors anymore until the end of the day \cite{announcement}.\\
Around 16:44 & Some people climb a pole and the narrow staircase next to the ramp (see Fig. \ref{fArea}). Several people try to elevate themselves from the crowd by climbing a billboard. Many seem to be in trouble between the staircase and the tunnel \cite{coolwojtek2}.\\
16:47 & Interview with the Love Parade organizer, who does not seem to be aware how critical the situation is \cite{Schaller}.\\
Around 16:48 & A command is given to stop inflows to the tunnel and the ramp area completely. It is executed within minutes \cite{Spiegel}.
Sound of police sirens; some people have fallen to the ground and raise their hands into the air for help \cite{coolwojtek3}.\\
Around 16:50 &  An emergency vehicle is entering the ramp area through the tunnel and opens its sliding door. An interaction between the crowd and people in the emergency vehicle takes place. The trouble between the staircase and tunnel is becoming more and more serious \cite{coolwojtek4}. 
A video from the West looking down on the crowd shows shockwaves in the crowd. Police forces are having a hard time holding a fence back at the container, which is used by the crowd management \cite{italian}. \\ 
\hline
\end{tabular}
\caption{Further relevant events.}
\label{SECOND}
\end{table}}
{\footnotesize\begin{table}[htbp]
\begin{tabular}{p{1.5cm}p{9.7cm}}
\hline
Starting 16:53 & The emergency vehicle stops in the middle of the crowd. 
Strong shock waves occur all over the crowd and push people to the ground between tunnel and staircase \cite{coolwojtek5}. Arms are lifted up and people are screaming. A group of people is aggressively pushing their way towards the tunnel (see Ref. \cite{coolwojtek5} between minutes 1:28 and 1:35). Some people are crawling on top of others to get towards the staircase. A helicopter flies overhead. Someone fixes a rope above the tunnel to allow people to climb up \cite{coolwojtek5,crawl}. \\
16:54-17:03 & Some people get pulled up to the narrow staircase. A ladder is lowered down to the container at the South end of the ramp, and   a woman, who seems to be hurt, is lying down on the container \cite{ladder,breezer1}.\\ 	
Starting 16:57 & People are pulled up one by one via the container \cite{cont}. 
People in the crowd are being pushed around. A few people climb onto other people, trying to get out of the crowd. A woman is screaming loudly \cite{screaming}. \\	
Starting 16:58 & The situation is extremely crowded. Some people scramble up the narrow staircase. Many people yell for help \cite{yell}.\\ 
Starting 16:59 & More people are pulled up from the crowded container to the festival area above. Security guards and police walk along the East side. A police officer is filming \cite{kaydee4}. An ambulance car is approaching on the freeway in the West.\\ 	 	
Starting 17:01 & View of emergency forces near the staircase area \cite{kaydee5}.\\	
Starting 17:02 & People scramble up the stairs. Many people are yelling for help. The situation is extremely crowded.
Police attempt to control the crowd \cite{artofhell2}. \\
17:02 & First victims are reported on the ramp \cite{wordpress,Polizeibericht}.\\
Starting 17:03 & The stairs are clearing slightly, and some people are able to get up \cite{artofhell3}.\\ 
Starting 17:03 & A man is trying to grab people and pull them up on the South over the container. Police holds the fence back.
An orange ladder is used to get people out from the container \cite{uwedix}.\\ 	
Around 17:04 & Seven policemen are talking to a few people. Two are helping someone on the ground \cite{kaydee6}.\\ 	
Starting 17:05 & A view from the tunnel shows some people climbing up over the container, also with the help of ropes. 
It seems that people in the tunnel behind are still reasonably fine. Some of them appear to be dancing
\cite{mbreezer8}. \\
Starting 17:05 & More people are able to get up via the staircase. The density in the ramp area is reduced, and the police are turning around some people at the back of the crowd, who are still trying to get to the stairs.\cite{artofhell4}.\\ 	
Starting 17:05 & A crowd of people has fallen in front of the stairs, raising their arms up. Some rescue workers and festival attendees are pulling people out. One policeman tries to hold back the crowd. An emergency vehicle is guided to the ramp area by the police, coming from the East tunnel \cite{kaydee7}.\\ 	
Starting 17:07 & The stairs are still crowded. Someone is shouting for help by the police. Some policemen on the stairs help people up \cite{artofhell5}.\\
Starting 17:08 & Someone is yelling at the police \cite{artofhell6}. People are pulled out of the fallen crowd, and some receive first aid. 
The crowd below the staircase seems ``cleared'' by the end of the video, and there is a considerable amount of police and rescue forces \cite{kaydee8}.\\ 	
Starting 17:08 & People can be seen lying on top of each other. The situation is still crowded, but the density eventually reduces \cite{rkjorge1}. \\
Starting 17:09 & The situation continues to be crowded, but people are starting to move more smoothly up the stairs. The area around the fallen people empties \cite{artofhellmore}. \\ 
17:15 & The operation room of the city of Duisburg does not seem to be aware of the critical situation. It still calls the Love Parade a big success \cite{success}. \\
Starting 17:16 &  The situation on the ramp has cleared up, but the group of fallen people still seems to be without professional help. A rescue crew appears in the South-West corner. A person is lying unconsciously on the ground. Many people try to resuscitate others. Fallen visitors are pulled out of the pile of people \cite{rkjorge2}.\\
Around 17:20 & The crowd has mostly dissolved. Fire and ambulance cars are parked in the South of the ramp. A woman tries to provide first aid to a man in the South-West corner. At least 2 other people provide first aid to people on the ground \cite{rkjorge3}. \\ 
Around 18:00 & It is decided not to terminate the Love Parade to avoid further critical situations (by evacuating the festival area too quickly) \cite{wordpress,weiter,Polizeibericht}.\\
\hline
\end{tabular}
\caption{Further relevant events (continued).}
\label{LAST}
\end{table}}
\clearpage
An overview of the videos (as well as the locations and times when they were taken) are provided on a supplementary webpage \cite{videocoll}.  However, we would like to point out that the times provided on the videos or in the respective video portal may not always be exact. A synchronized video collection is now also available  \cite{Videos}.

\section{Contributing Factors}

After the occurrence of a disaster, it is natural to ask, who is responsible. In many cases, 
people are trying to find one person or organization (the `scapegoat') to blame. In fact, after the Love 
Parade disaster, it seems that everybody was blaming everybody else: the visitors, the organizers, the police,
the city of Duisburg. What makes things difficult is that nobody is totally right and nobody is totally wrong: in the following, 
we will argue that it is the interaction of many contributing factors that caused the crowd disaster. Before we discuss
the interaction of these factors, however, let us shed more light on some of them in separation. While doing so, we will
address a number of hypotheses regarding the cause of the crowd disaster, which have been formulated after the event.
Given the many victims and pictures reminding of a war zone, some people first thought that a terrorist attack with explosives had happened 
\cite{RTLIIPart1}. Others claimed that the fatalities resulted, because some people had fallen on top of others when unsuccessfully
trying to climb the stairs from the side or the billboard \cite{stairfalling} (see Fig. \ref{fArea}). And again others were blaming the crowd for the outbreak
of a `mass panic' (stampede) \cite{PanicCause} or at least some people for improper behavior \cite{WrongBehavior}. 
The first hypothesis was obviously not true. But what about the others? 

\subsection{Did the crowd panic?}\label{PaniC}

When talking about crowd disasters, public media often use the term `mass panic', which suggests the occurrence of a stampede
as reason of the disaster (see Ref. \cite{asphyxia} and also the name of the link in Ref. \cite{Wiki}). This suggests that crowd disasters happen, because  the crowd `goes mad' \cite{MadnessOfCrowds}. There certainly exist some instances of this kind (such as the stampede
in Baghdad on August 25, 2005, due to spreading rumors of an imminent suicide bombing in the crowd \cite{Bagdad}, or the stampede in a Chicago night club triggered by rumors of a poisonous gas attack \cite{Chicago}). However, the hypothesis of a ``psychological state of panic'' as reason of crowd disasters has been questioned many times \cite{nopanic1,nopanic2}. 
\par
What evidence do we have for the Love Parade disaster in Duisburg? Has the crowd `gone mad' because of
influence of alcohol and drugs or because of impatience to get on the festival area? At first sight, one may think
so, given that a number of visitors climbed over fences, up the pole, and on the container to
reach the festival area. However, as we will see, these activities started at a time when people on the ramp were already exposed
to crowded conditions. 
\par
Let us discuss this in more detail. The first problems with visitors overcoming fences were reported around 15:31 \cite{Boeschung}. However, there were reports as early as 13:40 (see Table \ref{SECOND}), which show that people waiting for access had difficulties to breathe and asked to open the emergency exits (which did not happen) \cite{EarlyProblems}. 
These problems demonstrate that the access capacity was far below demand. 
\par
Problems related to queues of people aggravate when queues are long and broad, so that little or no progress is visible. In such situations, 
people will subconsciously reduce their distance eventually. Although the reduction of distance might be negligible, the so-called `queuing effect' will create the impression of progress. However, it will also cause a compression of the crowd \cite{queuing}. When the distance is small, there will be inadvertent body contacts, which can add up and cause unintentional pushing. Note that the transition from an acceptable situation with rare body contacts to a stressful situation with frequent contacts can happen quite abruptly  \cite{turbulence}. People may interpret this as intentional pushing, which may trigger stress and aggression. At a certain density, it may also be required to push others away in order to be able to breathe \cite{WenjianAnders,EarlyProblems}.  
\par
If people have to wait long and are not informed about the reasons for this, they will become impatient and may eventually start to push {\it intentionally} (because they assume that progress can be accelerated). While most impatient pushing happens in the middle of the queue, the situation usually becomes most critical at the front of the queue (but the people who push cannot see this, and they experience much less crowded conditions). 
\par
The situation is particularly bad behind bottlenecks. These can create `traps' without any possibility to escape. Such situations must generally be avoided. This also means that flow control is not a solution for every problem. It requires suitable designs and an adaptive operation. 
\par
According to our assessment, it had to be expected that the access points would have to be opened and fences would eventually be overcome, given that the festival area and the inflow capacity were small (in particular as the access was delayed by leveling works). Waiting times often amounted to several hours, and access to entertainment, food, water, and toilets must have been quite limited outside the festival area. 
\par
Nevertheless, the problems on the ramp were even more serious than at the access points. They were related to the low inflow to the festival area (see Table \ref{flows}). An analysis of surveillance videos suggests that the floats (i.e. the moving music trucks) `pulled' visitors along with them, as expected by the planners, but this was not apparently effective enough. After the crowd disaster, it was sometimes claimed that the floats even obstructed the inflow of arriving visitors. While the inflow never stopped completely before the cordons were established \cite{inflow}, the queue forming at the top of the ramp varied considerably over time \cite{inflow,Rampe,floats}. The inflow was particularly low, when a float was slowed down or stopped around 15:31 in the neighborhood of the ramp \cite{FLoaT}. 
\par
While the organizers considered the possibility of inflow problems \cite{Entfluchtung,Schaller}, they assumed that these could be handled by `pushers'\footnote[1]{`Pushers' are people, who are supposed to put pressure on visitors to keep moving, in this case to ensure an efficient entering into the festival area in order to avoid an obstruction of other visitors trying to get in.} 
at the upper end of the ramp and that the floats could be used as well to {\it reduce} them (by attracting the crowd onto the festival area and moving it along with them) \cite{Entfluchtung}. However, there was apparently a lack of a sufficient number of pushers \cite{Spiegel}, and the floats did not manage to overcome the inflow problem. It looks like the floats were slowed down by the dense crowd, which in turn obstructed the inflow of visitors, thereby creating an unfavorable feedback loop.
The situation was particularly tense from 14:27 to 15:05 and from 15:55 to 17:00; as a consequence, the crowd manager asked for support by the police at 15:16 (or before) \cite{Polizeibericht}. The responsible officer arrived around 15:30, when a jam had formed on the upper part of the ramp \cite{kaydee271}. About 10 minutes later, a joint strategy was found. However, already at 15:31 (i.e. at the time when one of the floats slowed down in front of the ramp), the situation had deteriorated so much that a large amount of visitors decided to overcome fences along the ramp to reach the festival area via the grassy slopes on both sides (see Fig. \ref{fArea}) \cite{Boeschung,inflow,Rampe}. This mitigated the bottleneck situation at the end of the ramp, which could have caused serious problems at a much earlier time. In fact, it seems that the dangerous phenomenon of crowd turbulence (see Sec. \ref{CrowdTurb}) first occurred in the upper part of the ramp \cite{floats}.
\par
According to Table \ref{SECOND}, the first visitors used the narrow staircase at 16:17, and around 16:22 the first people climbed the pole on the East side of the lower ramp area, to get up to the festival area \cite{Pizzamanne}. The first people climbed the container of the crowd management on the South of the ramp at 16:24 \cite{containerclimb}. This was the time, when the third police cordon is given up. While the initial flow on the staircase was stopped by police, people used the staircase again around 16:27. At about the same time, a person climbed a traffic sign on the ramp \cite{signclimb} (see Fig. \ref{fArea}). 
All of this might have been interpreted by the security as signs of an excited crowd that did not behave properly, but the temporal coincidence of these events clearly shows that people were trying to escape from the crowd in any possible way, because they felt in danger. In fact, behavior of the crowd that might have been perceived as `improper' occurred mainly, after the first two cordons had to be given up (around 16:14 and 16:21), while the third one was still closed, which caused an increasingly crowded situation on the ramp. 
\par
In videos recorded at the Love Parade, the phenomenon of crowd turbulence starts to appear between 16:34 and 16:36 \cite{lightpole}. Around the same time one can hear painful shouts, and some people scream for their lives and for help \cite{scream} (see Table \ref{LAST}). In this situation, at least some people must have experienced a psychological state of panic. Nevertheless, there were no signs of sudden systematic movements of the crowd into a certain direction, which would indicate a stampede, and no people `crawled' on top of others, yet. Around 16:40, the forces in the crowd were so critical that a traffic sign was bent \cite{moveon}, and an unconscious women was passed on to the narrow staircase \cite{unconscious}. Around 16:45, several people tried to elevate themselves out of the crowd by climbing a billboard next to the traffic sign \cite{coolwojtek2}.  Approximately at the same time,  many people between the billboard and the staircase raised their arms into the air \cite{coolwojtek3} (the movie should be watched in full screen mode to see this well). This is usually a sign that they have fallen to the ground and are seeking help from others to get back on their feet. We believe that this was the first sign that people were dying or likely to die. At 16:51, an emergency vehicle entered the ramp, but it was taking care of other problems \cite{Pizzamanne}. Still, there were no sudden moves into one direction visible in the crowd that would speak for a stampede. Rather, people next to those screaming for their lives were trying to calm them down by saying ``you will make it,'' and offering them water \cite{helping}. Around 16:55, a group of people was pushing their way through the crowd towards the tunnel in the West (in Ref. \cite{coolwojtek5} this can be seen between 1:28 and 1:35 in full screen mode; the same shows up in Ref. \cite{ADDitional}). Around the same time, some people were trying to `crawl' over others, hoping to escape the situation \cite{coolwojtek5,crawl}. While this was clearly a relentless and potentially harmful behavior, it is not obvious that it killed others, and it occurred under circumstances that were absolutely life-threatening (which should not be misunderstood as a justification of such relentless behavior.) First deaths were reported at 17:02 \cite{wordpress,Polizeibericht}.

\subsection{Were people killed by others falling on them from above?} \label{Stairs1}

As most people died between the staircase and the billboard \cite{wordpress}, the public media initially assumed that they were victims of others, who had fallen down after unsuccessfully trying to climb the staircase from the side or to climb the billboard \cite{stairfalling}. There was even a statement that the staircase should have been ``blasted away'' before the event \cite{BlastAway}. However, the videos viewed by us do not provide convincing evidence that falling people were the cause of the disaster. It is also not plausible that a few people falling from the staircase could account for 21 fatalities and more than 500 injured people \cite{Wiki}. Moreover, the height of falling was not large, 
and most victims were not lying on the side of the staircase, but rather between the staircase and the entrance of the tunnel \cite{FOTO} (see ``accident area'' in Fig. \ref{fArea}).
\par
Nevertheless, the analysis of the video materials and photographs witnesses at least three events of falling people. According to Ref. \cite{wordpress}, the first one happened around 16:57 at the billboard, the second shortly later at the same place. The third incident happened at the same location at 17:03. Furthermore, one person failed to climb the staircase from the side; around 16:40 it fell back to the ground from a low elevation \cite{FallingFromStairs}. Apparently, the height of falling was relatively small, and the falling people also did not trigger a stampede of the crowd. Therefore, according to our judgment, it is unlikely that people died as a direct consequence of others falling down from the staircase or billboard. 

\subsection{Did the Staircase Cause a Crowd Crush?}

Nevertheless, it is a valid question, whether it was a mistake to let people use the staircase. It is likely that people were turning towards the staircase, hoping that it would provide a chance to escape, and that even a minor movement could seriously increase the local pressure in the crowd, given the high density that had already built up on the ramp. In fact, the situation in the crowd was highly problematic not only next to the staircase \cite{pizzamanne}, but also next to the pole \cite{lightpole}, and it was apparently the use of the pole that inspired the use of the staircase \cite{WirNehmenDieTreppe}. Nevertheless, the movement of the crowd towards these improvised `emergency exits' was not large. The videos we have seen do not show the sudden start waves, which are typical when a waiting crowd (or jammed traffic) starts moving \cite{queuing,RMP}. Therefore, we doubt that the fatalities were caused just by a relentlessly forward pushing crowd, which crushed the people. Crushing due to extreme densities rather happens when a large crowd moves too quickly towards a narrowing \cite{sim4}. In Duisburg, however, the crowd disaster happened in a crowd that barely moved forward. Even though the situation on the ramp was critical for the crowd from 16:35 on \cite{Pizzamanne}, it seems that most people had a chance to breathe (at least intermittently) and to recover between stressful periods. In fact, the recordings change many times between screams of panic and more positive noises. 
\par
We do not question that the density in the crowd became so high at some locations that it could seriously harm health and lives, but it is puzzling that most victims were not found on the side of the staircase, or next to the pole(s) and the container, where they had to be expected in case of a crowd crush. We also do not deny that the staircase was an attraction point, but we doubt that it can be seen as immediate cause of the disaster. It may have even played a significant role for the evacuation of the overcrowded ramp, since it served as emergency exit. However, this emergency exit was used too late and not very efficiently. A continuous flow of people on the staircase was established only around 16:40 \cite{helping, Pizzamanne}. Before, it stopped or was blocked many times. The same happened during the most critical period, when many people tried to climb the staircase from the side, which considerably obstructed the flow on it \cite{ClimbFlow}. 

\subsection{Occurrence of Crowd Turbulence} \label{CrowdTurb}

So far, the cause of the crowd disaster in Duisburg has still not been revealed.
If the crowd did not panic, and people did not die from others falling on them, and a rush towards the narrow staircase did not cause the crowd disaster, what then was the reason for it? The answer lies in the dynamics of the crowd, which unintentionally emerged, when the density became too high. 
John Fruin describes the situation as follows \cite{Fruin}:
``At occupancies of about 7 persons per square meter
the crowd becomes almost a fluid mass. Shock waves
can be propagated through the mass, sufficient to  ...
propel them distances of 3 meters or more... . People
may be literally lifted out of their shoes, and have clothing
torn off. Intense crowd pressures, exacerbated by
anxiety, make it difficult to breathe, which may finally
cause compressive asphyxia. The heat and the thermal
insulation of surrounding bodies cause some to be
weakened and faint. Access to those who fall is impossible.
Removal of those in distress can only be accomplished
by lifting them up and passing them overhead
to the exterior of the crowd.''
\par
In fact, suffocation was diagnosed as the reason for the death of
people during the Love Parade disaster \cite{Wikipedia}. In simple words, this means that the lungs of the victims
have been compressed so much that they were unable to breathe enough to get the required
amount of oxygen to survive. Compressive asphyxia was also identified as cause of death in 
many other crowd disasters. 
\par
According to recent studies \cite{turbulence}, it is often not the density alone
that kills (`crushes') people, but the particular kind of dynamics that occurs when the density is so high that 
physical interaction between people inadvertently transfer forces from one body to another. Under such conditions, forces in the crowd can add up.
Force chains may form, such that the directions and strengths of the forces acting on the body of an individual
in the crowd are largely varying and hard to predict. As a consequence, an uncontrollable kind of collective dynamics 
occurs in the crowd, which is called `crowd turbulence' or `crowd quake' \cite{turbulence,crowdsafety1}. The forces in this dynamical state of
the crowd can cause various injuries (in particular of the chest, as in crowd crushes). They are so high that they cannot even be controlled by large numbers of police forces. Individuals can handle the situation even less. They are exposed to a large risk of losing balance and stumbling \cite{stumbling}. 
\par
Once people have fallen, they constitute obstacles to others and are endangered by others falling on top of them, since these
can also not control their steps anymore as they wish. Hence, the surrounding people are likely to stumble as well, which creates a `domino effect' \cite{Domino}.
The resulting number of falling people may be large. This creates a heap of people, in which nobody can easily get back
on their feet again. Those on the bottom have serious difficulties to breathe, and they are likely to suffocate if this state lasts too long, given the weight of others on their top. 
\par
\begin{figure}[htbp]
\begin{center}
\includegraphics[width=8cm]{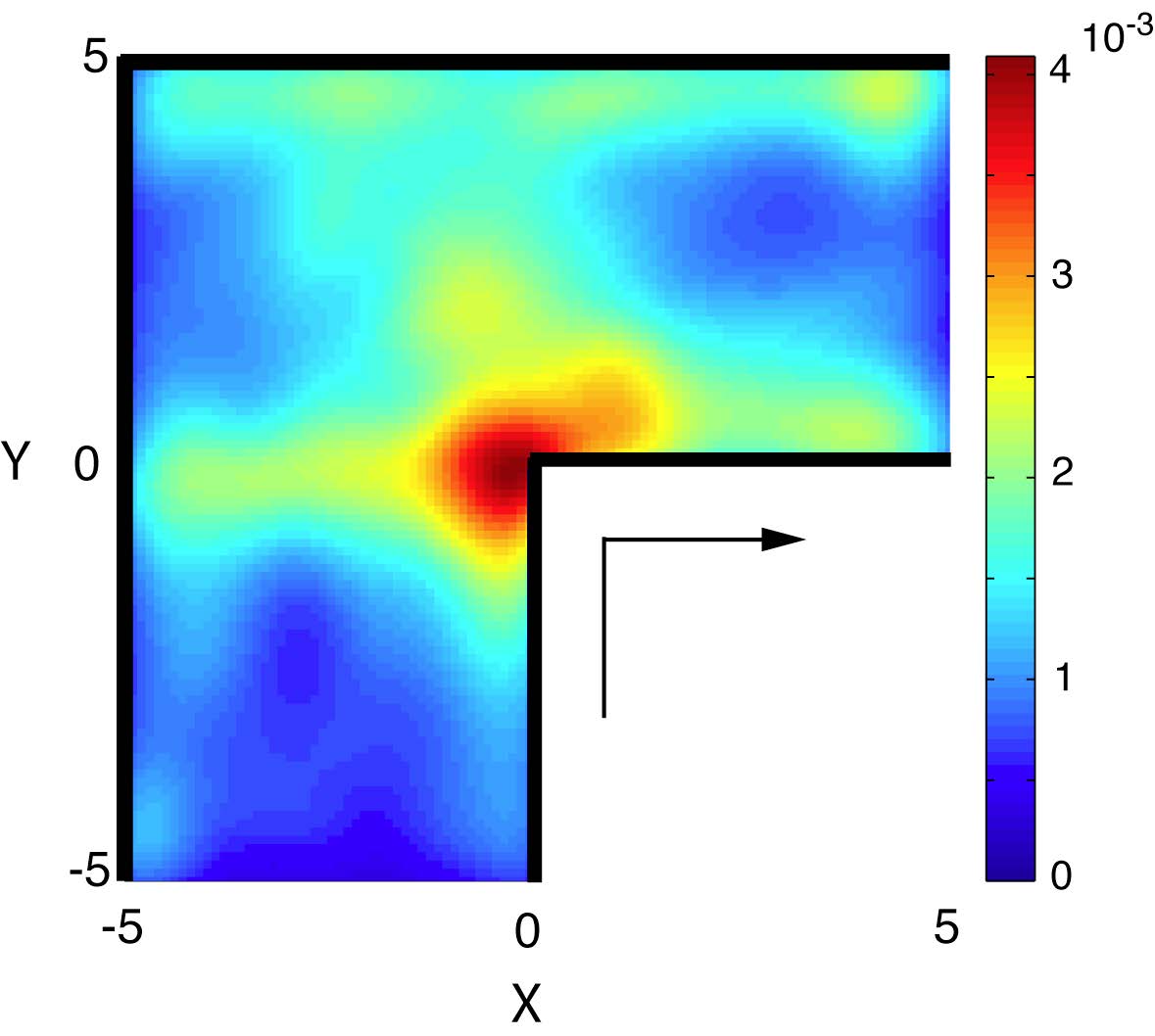}
\end{center}
\caption[]{Computer simulation of a densely crowded area with the heuristic pedestrian model of Ref. \cite{sim2} (figure from the Supporting Information). Orange and red areas indicate the locations with the highest crowd pressure, when a densely packed crowd tries to move around the corner. The simulated situation is analogous to leaving visitor streams at the Love Parade in Duisburg, trying to leave the main ramp through the tunnel in the West. Note, however, that the above simulation does not consider inflows of visitors arriving from the East. These would move the high-pressure area a bit up the ramp, where the accident actually happened.}
\label{simulat}
\end{figure}
Directly after the Love Parade disaster, when the situation was far from clear,
one of the authors conjectured that `crowd turbulence' was the likely cause of the fatalities \cite{turbulenceInNews}. Eye witness reports \cite{eyewitness} and the analysis of video recordings confirms this hypothesis. Crowd turbulence can be observed in the crowd at least from about 16:34 on around the pole and from 16:39 on in the lower part of the ramp \cite{TURbu}. 
Before 16:48, a considerable number of people fell to the ground between the tunnel and the staircase \cite{coolwojtek3}, approximately at locations where computer simulations predict the largest crowd pressures (see Fig. \ref{simulat}).\footnote[2]{Note that the fallen people in the video recording are in the shadow. Therefore, one must use full screen mode to notice them, and one needs to watch out for arms raised in the air, seeking for help.} The situation deteriorated further around 16:53, when crowd turbulence affected almost the entire width of the ramp \cite{coolwojtek5}, i.e. hundreds or even thousands of people were irregularly moved around by the pressure in the crowd; many of them stumbled and fell on top of each other \cite{coolwojtek5}. The troubled area agrees with the one, where most victims were found \cite{LyingOnGround,FOTO}. Under the weight of others lying on them, they must have eventually suffocated, since there were not enough emergency forces to help them back on their feet in time. 
\par
Public blogs have been wondering about the reasons for the layered crowd of fallen people \cite{wonder}:
\begin{enumerate}
\item Did the emergency vehicle driving on the densely crowded ramp trigger the falling?
\item Was there a fence lying on the ramp, that should have covered a broken manhole cover \cite{manhole}? 
\end{enumerate}
While cars moving through a dense crowd can indeed trigger critical conditions, it seems that people had already fallen to the ground (around 16:48), before the emergency vehicle arrived on the ramp (around 16:50 \cite{Pizzamanne}).\footnote[3]{Moreover, video recordings of the situation around the emergency vehicle do not show clear evidence of turbulent motion in its immediate neighborhood \cite{noturb}.}
\par
A broken manhole cover or any kind of obstacle lying on the ground would certainly have made it difficult for people to keep their balance and stay on their feet, when pushed around by turbulent waves. Such obstacles are dangerous and should certainly not have been located in the bottleneck area (the ramp). While, even without obstacles, it is likely that crowd turbulence would have caused people to fall sooner or later, obstacles can act as `nucleation points' and thereby possibly trigger an earlier falling of people, which may reduce their chances of survival.

\section{Causal Interdependencies}

We must now discuss the question, how the conditions, which caused the deadly crowd turbulence, have come about.

\subsection{Failure of Flow Control}

When viewing the area of the Love Parade in Duisburg (see Fig. \ref{fArea}), the choice of location appears surprising, since the festival area was relatively small
and furthermore constrained by railway tracks on one side (in the East) and by a freeway on the other side (in the West). This becomes particularly clear when comparing the area with the one used during the Love Parades in Berlin (see Ref. \cite{comparison}). As this circumstance implied a risk
and the bottleneck at the ramp during peak hours was foreseeable (see Sec. \ref{festa}), flow control was crucial for the safety of
the Love Parade. However, there was a whole avalanche of problems that accumulated and, thereby, caused the crowd disaster.
\par
The first problem on the day of the Love Parade occurred when the opening of the festival had to be delayed by approximately one hour due to a delay in the completion of the leveling work (see Table \ref{SECOND}). Therefore, many visitors must have been queued up already at the time when the festival area was opened. It seems that the organization of the mass event could never make up for this delay. 
\par
The overall inflow capacity was apparently further reduced through obstructions by the floats, which had probably not been anticipated to that extent (see Sec. \ref{PaniC}). As a consequence of this, access control was necessary already at 13:00 (see Table \ref{FIRST}), much before the expected peak hours. This further increased the queues and the waiting times. The following quote witnesses the problems \cite{PrivComm}: ``We parked the car about 3 kilometers away from the  freight station (next to the festival area), and it took us almost 5 hours (!) to get to the Love Parade (festival area). On the way, we were facing blocked roads time and again, fences were carried over us, emergency forces could not get through, people collapsed, ...''
\par
Clearly, visitors of the event must have become impatient, particularly because there was probably a lack of food, drinks and toilets outside of the festival area (since such long waiting times were not anticipated). One could, therefore, expect that it would be difficult to control the inflow. In this connection, it is also worth noting that there was not much entertainment outside the festival area to shorten the psychological waiting time and to relieve stress and impatience. Apparently, there was a stage outside the festival area, which was supposed to absorb some of the visitors, who could not get to the festival area, but for some obscure reason, it was moved to another area, where it attracted only a smaller number of people \cite{Spiegel}.
\par
Between 14:30 and 15:10, the organizers found it difficult to control the inflow with the isolating devices (see Table \ref{FIRST}). This was probably not just a result of the excessive waiting times, which caused impatience, but possibly also because some of their security people were needed elsewhere (e.g. to improve the outflow from the ramp or to guide VIPs) \cite{Spiegel}. As a consequence, the organizers tried to get support by the police \cite{Spiegel}.
\par
For a number of reasons, it seems to have taken a considerable amount of time to get the requested police support. Communication by walkie talkies and mobile phones did not work reliably \cite{Spiegel}. There were also no functioning loud speakers at the ramp, as there should have been \cite{Spiegel}. Moreover, there was a change of police shifts between 15:12 and 15:34, when the situation started to deteriorate  \cite{Spiegel}. Various reports suggest that police and organizers were not well coordinated, probably due to the afore-mentioned communication problems. It is also likely that the following emergency operations had not been exercised before. As a consequence, the police may have tried to solve the problem with concepts they were familiar with. They formed several police cordons for flow control. This tactic is often applied to get control of violent crowds. However, it failed during the Love Parade, and we will now analyze why.

\subsection{A Lack of Overview of Everybody}\label{LACK}

As was pointed out in Sec. \ref{KStill}, when the crowd was trapped in a situation of extreme density, it did not have a chance to get an overview of the situation and possible ways to improve it, in particular to get out of the area. Signs and loudspeaker announcements were not available. The only possible emergency exits they could recognize were the narrow staircase, the pole(s), and the container of the crowd management. They were used accordingly, which was quite reasonable in the more and more dangerous situation that the crowd found itself in. 
\par
At this time, all the hope to get control of the situation rested on the police. The police may have been surprised by the sudden need to take control, which was requested by the crowd manager when difficulties to access the festival area occurred at the upper end of the ramp. 
The police tried to solve the problem by establishing cordons, but it was soon noticed that police cordons 1a, 2, and 3 (see Fig. \ref{fArea}) blocked not only the inflow, but the outflow as well. This is also the reason why cordon 1a was moved behind the side ramp (see cordon 1b in Fig. \ref{fArea}), and why police cordon 4 was formed at the upper end of the ramp (after dissolving cordon 3). This would have allowed to re-direct the outflow via the side ramp. However, before these operations could be completed, cordons 1b and 2 had to be given up because of the increasing pressure in the waiting crowd, while cordon 3 was still there \cite{Polizeibericht}. 
\par
It is known that dense counter-flows are unstable and may give rise to mutual blockages, which can cause crowd disasters \cite{evacuation,counterflows}. For such reasons, it is recommended to separate the flow directions at mass events. Yet, it was not the instability of dense counter-flows which caused the incident in Duisburg. The lack of directional flow separation, however, did not allow one to clear the ramp, after it became crowded by the dissolution of two of the cordons. When cordons 1b and 2 had to be given up, the police suddenly found itself in a situation, where in- and outflows blocked each other, and it was basically impossible to evacuate the ramp in conventional ways, when people quickly accumulated on both sides of cordon 3. 
A trap without exits or emergency exits resulted, from which people could not get out, and the situation kept getting worse  \cite{Augenzeuge}. 
\par
For people in the crowd, it was impossible to gain a sufficient overview of the situation and to find a solution. Police had helicopter surveillance \cite{helicopter} and was filming the ramp from the top. However, it took some time until the criticality of the situation was noticed and evacuation measures were taken. When the evacuation finally became effective, the ramp cleared quickly \cite{rkjorge3}. But prompt action was delayed by communication problems. It seems that the first loudspeaker announcement could only be made around 17:30, after a loudspeaker vehicle had entered the ramp \cite{late}. 
\par{\footnotesize\begin{table}[htbp]
\begin{tabular}{p{1.5cm}p{9.7cm}}
\hline
14:27-15:05,
15:55-17:00 & Queues of arriving visitors form at the upper end of the main ramp, which leads to the festival area. For this case it was planned (1) to use `pushers' in order to make the people move forward, (2) to close the access points in the East and West in front of the tunnels, (3) to make loudspeaker announcements [pp. 20+13]. \\
15:16 & The crowd manager asks for police support via the liaison officer [p. 31].  \\
Around 15:30 & The relevant police officer arrives at the container of the crowd manager [p. 31]. \\
15:30-15:40 & Crowd manager and this police officer jointly decide (1) to ask crowd management/security staff to work as `pushers' in order to ensure a better inflow into the festival area from the upper end of the ramp, (2) to close the access points for approximately 10 minutes, (3) to form a cordon in the middle of the ramp in order to shield visitors trying to enter the festival area from behind. [pp. 20+31]\\
15:45 & In the discussion with other police officers, this plan is modified towards forming 2 police cordons in the tunnels to the West and to the East [p. 22]. \\  
15:50-16:20 & Police cordon 1 is formed in the tunnel in the West (first before the side ramp and then after it from 16:02 on in order to allow people to use the side ramp) [p. 21]. \\
15:57-16:16 & Police cordon 2 is formed in the tunnel in the East [p. 21]. \\
16:01-16:24 & A third police cordon is formed in the middle of the ramp in order to avoid that visitor flows returning from the Love Parade would undermine police cordons 1 and 2 from behind [p. 21+22]. \\
Around 16:10 & When arriving at the relevant area of the ramp, the responsible officer discovers that (1) many people are trying to leave the festival area and (2) 
the expected dissolution of the jam at the upper end of the ramp did not happen within the 10 minute time period foreseen for this. Therefore, the blockage of the inflows by cordons 1 and 2 must be maintained longer than planned. Due to this delay and since the access points must be intermittently opened, the pressure on police cordons 1 and 2 becomes so high that they must be given up [p. 23]. \\
16:24 & Visitors are jammed up on both sides of police cordon 3. The situation becomes extremely crowded [p. 24]. Therefore, 
police cordon 3 is dissolved, also because it is ``ineffective'' between two oppositely directed flows [pp. 24+34]. \\
16:31 & A new (transparent) police cordon is formed at the upper end of the ramp from 16:31 on [pp. 21+24]. It serves to stop the outflow of leaving visitors via the main ramp and to encourage arriving visitors to use the slopes to enter the festival area (see Fig. \ref{fArea}).\footnotemark[4] 
[pp. 24+34]  \\  
16:39 & The fire brigade reports `panic-like' movements on the ramp with some over-run people [p. 25]. \\
16:40-16:55 & The festival area is closed for newly arriving visitors (by moving vehicles in front of the access points) [pp. 25+35]. \\
After evacuation of ramp area & Some densely crowded spots remain around the container, two poles and the narrow staircase. It is not possible to redirect them by words or gestures [pp. 34+35]  \\
\hline
\end{tabular}
\caption{Course of events as presented in the police report \cite{Polizeibericht}. The numbers in square brackets correspond to the page numbers of the report.}
\label{POL}
\end{table}}
Why did the evacuation start so late? The analysis of the police is presented in Table \ref{POL}. It seems that first attempts to direct the crowd towards the upper end of the ramp started around 16:40 \cite{wordpress,firstattempts}, but were not very effective \cite{Polizeibericht}. It is true that evacuation attempts take some time, but there was also a lack of efficient means of communication (such as loudspeakers or megaphones). Moreover, we would like to point out the following: In crisis situations, decision-makers are often overwhelmed by the pace of events \cite{Physica_A}, mainly for two reasons: First, it takes time to collect information locally, and bring it to the attention of the chief police officer, who then takes a decision and gives commands. These are then transmitted down to the local police forces through the command chain. Second, critical situations are often characterized by incomplete, contradictory, and ambiguous information, which makes it difficult to assess the situation correctly and come to the right conclusions. 
\par
When the situation on the ramp became unbearable and life-threatening, people started to escape via the pole, the container and the staircase next to the ramp. This could have been misinterpreted as aggressive attempts of impatient visitors to storm the festival area, but in reality, it was a sign of emergency. However, due to the noise level, screams for help \cite{screaming,yell} were hard to comprehend. Also visitors (on the East), looking on the ramp from above around 16:30 did not have a sense of emergency \cite{NoSenseOfEmergency}. This makes it understandable, why pressure relief operations were not yet effective, when the crowd disaster was about to start. 
\par
Once the evacuation process on the ramp started, the area emptied quickly \cite{rkjorge3}.\footnotetext[4]{This was previously prevented by fences.}  The narrow staircase also might have played an important role as an emergency exit at this time \cite{helping}. Others managed to leave the ramp towards the festival area, following the emergency vehicle \cite{FollowPolice}. However, people close to the staircase were still focused on it \cite{WirNehmenDieTreppe}. This might have been a result of the `tunnel vision' that develops when people are stressed. Even when the surrounding crowd had dissolved, it took a long time, until those who had fallen to the ground between the tunnel in the West and the staircase got back on their feet, if they managed this at all \cite{coolwojtek5,LyingOnGround}. In fact, many of them were injured or died.
\par
A lack of overview is typical for crises situations. During the Love Parade disaster in Duisburg this is, for example, reflected by the fact that, around 15:06, the minister of interior visited the Love Parade (see Table \ref{SECOND}), but despite first signs of overcrowding, he left the festival area before the incident. At 16:47, the organizer gave an interview, which still called the event a success \cite{Schaller}, and as late as 17:15, the city's situation room made a similar statement \cite{success}. Emergency forces were also responding late. As a consequence, a triage procedure had to be applied. (This procedure is typical for war zones, major disasters, and terrorist attacks.) Therefore, many people in critical health conditions did not get first aid \cite{Triage}.

\section{Discussion}

In the following, we try to gain an integrative view of causal factors of the crowd disaster, which strictly needs to be distinguished from a legal analysis or a determination of responsibilities. We also want to stress that the main purpose of our analysis is to learn for the future, i.e. to identify factors that need to be paid more attention to. 

\subsection{Resilience, Systemic Instabilities, and Cascading Effects}

Note that, generally, a good organizational concept should be resilient (`forgiving'), i.e. it should be robust to mistakes and complications. Therefore, many disasters do not have a single causing factor. They are a result of interaction effects. This also applies to the Love Parade disaster which, as we will argue below, can be understood as result of a systemic instability.\footnote[5]{or even several interrelated systemic instabilities (since the phenomenon of `crowd turbulence' itself can be seen as outcome of an instability of visitor flows)}
The term `systemic instability' is used here for situations, where small perturbations can trigger a series of events through mutual amplification effects in a way that things eventually get out of control, even if everyone makes best efforts. At the Love Parade, people were dying although nobody wanted this and everyone was trying to prevent the death of people. Other examples for systemic instabilities are
\begin{itemize}
\item spontaneous breakdowns of traffic flows above a certain critical density (even when everyone is driving in a circle and trying hard to maintain a finite speed) \cite{RMP,trafficcircle},
\item breakdowns of cooperation in social dilemma situations, which give rise to `tragedies of the commons' \cite{Hardin},
\item political revolutions \cite{revolutions1,revolutions2,revolutions3} and
\item financial breakdowns \cite{IRGC,queen}.
\end{itemize}
Many systemic instabilities come along with cascading effects, which tend to create extreme events \cite{Spread1,Spread2,IRGC}: the overload of one component of the system challenges other components, which therefore causes a propagation of problems through the system. Usually, cascading effects do not occur during normal operation, but are triggered by (random) perturbations or the coincidence of several complicating factors. They tend to occur when the interdependencies in the system exceed a critical strength. For example, cascading effects are observed in traffic jam formation (when the density is too high) \cite{RMP}, blackouts of power grids, for many kinds of disasters \cite{BlackoutCascades}, for the current financial crisis \cite{IRGC}, and for the Arab Spring revolutions \cite{revolutions1,revolutions2,revolutions3}. 

\subsection{What Caused the Crowd Disaster: Causal Interdependencies of Contributing Factors}

\begin{figure}[htbp]
\includegraphics[width=12cm]{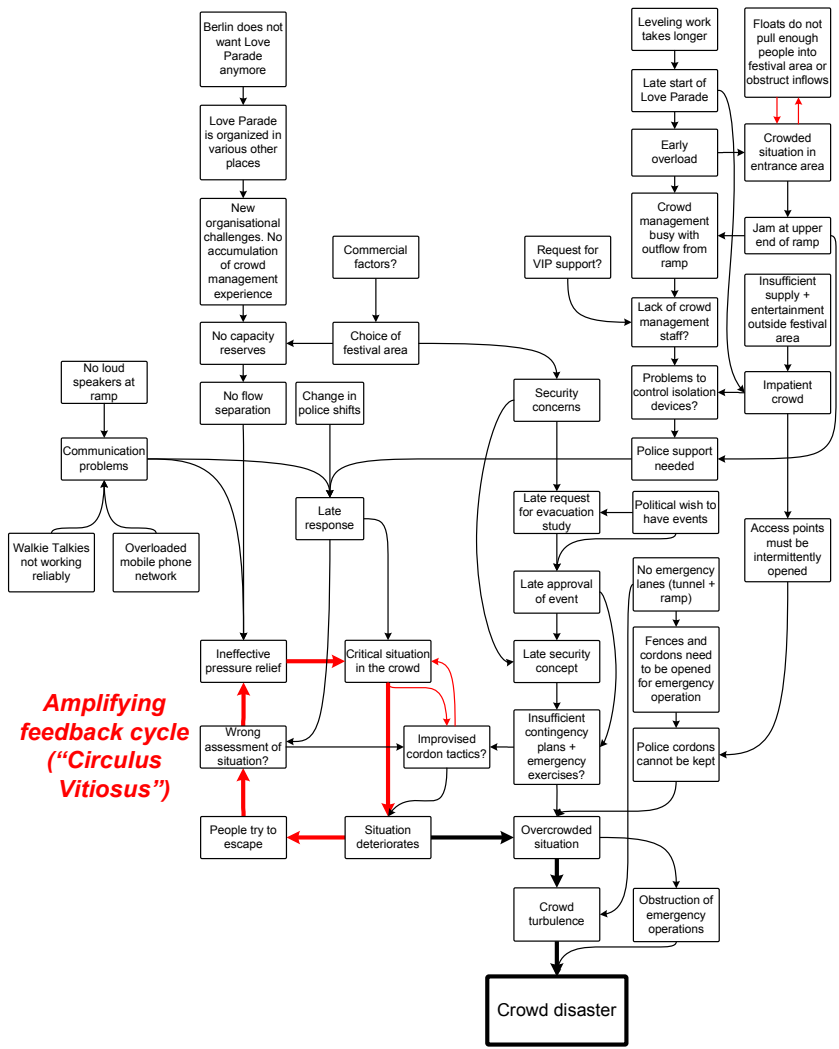}
\caption{Illustration of causal interdependencies between different factors that have most likely contributed to the emergence of the crowd disaster during the Love Parade in Duisburg. One can see that the reason for the crowd disaster was not a single factor, but amplifying feedback and cascading effects, as it is typical for systemic instabilities. Therefore, most contributing factors are consequences of other factors. Also note that causal dependencies have to be clearly distinguished from legal responsibilities. (Question marks indicate likely contributing factors, which we have not proven by us in a strict sense, but are plausible conclusions that are not questioned by any counter-evidence known to us.)}
\label{Cascading}
\end{figure}
The following analysis discusses cascading effects that have (most likely) contributed to the Love Parade disaster in Duisburg (see Fig. \ref{Cascading} for an illustration). 
\begin{itemize}
\item Berlin rejects to host the Love Parade (LP), and other cities take over \cite{Wikipedia}. The Love Parade moves from city to city, which creates new organizational challenges each time (in more difficult locations than in Berlin with its wide roads and expansion areas). The change of organizational teams makes it difficult to accumulate crowd management experience over many events. 
\item Bochum has to cancel its Love Parade, because it cannot manage the security challenges \cite{Wikipedia}.
\item Duisburg/Essen is elected as cultural metropole 2010 \cite{metropole}. It is under pressure to come up with an attractive cultural program. This seems to have created a desire to approve the Love Parade \cite{Spiegel,political}. 
\item The festival area does not provide capacity reserves and implies a number of organizational difficulties. In the tunnel and on the ramp, in- and outflows are not separated, and there is no separate route for emergency vehicles (i.e. they have to use the tunnel as well). 
\item To overcome security concerns, an evacuation study is commissioned. It mainly focuses on evacuation scenarios \cite{Entfluchtung},
assuming a maximum concurrent number of visitors as it was required by the security concept of the city \cite{Spiegel}.\footnote[6]{Some tolerable risks associated with the normal entering and leaving of the area are mentioned, but have not been investigated in detail by computer simulations.}  
\item Due to the late approval of the event (see Table \ref{SECOND}), the security concept may have been finished `last minute' (and vice versa). The likely consequence is that contingency plans may have been insufficient and could not be exercised enough. There was probably also not enough time to ensure a good coordination between organizers and police forces. 
\item Due to delays in finishing the leveling work (see Table \ref{SECOND}), the festival area of the Love Parade is opened later than expected \cite{LateStartNoSeparation}. This implies an early overload of the access points and causes an impatient crowd (particularly as facilities, supply and entertainment were probably scarce outside the festival area). 
\item People enter the Love Parade area later and return earlier than expected. 
\item The interaction of the floats with the crowd does not enable a sufficient inflow to the festival area. This apparently requires that crowd management forces are moved away from the isolating devices to the end of the ramp, in order to improve the inflow;  requested VIP support seems to absorb some manpower as well \cite{Spiegel}. 
\item The crowd management faces problems to control the isolation devices, and it tries to organize police support \cite{Spiegel}.
\item There are difficulties in the communication and coordination between organizers and police. Suitable communication means are missing or not used or are not working in a reliable way \cite{Spiegel}. Therefore, the feedback between the situation, the crowd management, and the crowd is insufficient. 
\item Due to communication problems and a change in police shifts, police support may have been delayed \cite{Spiegel}. Moreover, it must have been difficult for the new shift to get an overview of the situation. 
\item Maybe due to the urgency of the situation, it is decided to form two police cordons in the tunnels leading to the ramp. A third police cordon
is established in the middle of the ramp, where fences narrow down the diameter of the ramp. It shall prevent that leaving visitors undermine the police cordons in the tunnel from behind \cite{Polizeibericht}. 
\item The police cordons in the tunnel are given up, probably because of the high pressure of the arriving crowd. This replaces the problem at the upper end of the ramp by an even bigger problem in the middle of it: A lot of visitors are moving into the lower ramp area through the tunnel, while many others are waiting at the upper end to leave the event. As the third cordon blocks in- and outflows, jams of arriving and leaving visitors are quickly growing on both sides of cordon 3. The cordon is dissolved, because it is ineffective, and a new police cordon is formed at the upper end of the ramp. 
\item At this time, the situation in the crowd is already critical. The lack of separation of opposite flow directions makes it difficult to let people out without letting people in \cite{LateStartNoSeparation}. Therefore, it is impossible to evacuate the ramp efficiently.
\item People on the ramp try to escape the life-threatening situation over the staircases, the pole(s), and the container (see Sec. \ref{PaniC}). This may have been misinterpreted as a `mob' trying to force its way into the festival area, which needs to be controlled. Pressure relief efforts become effective only very late. 
\item In absence of separate emergency routes, fences and cordons must be opened to allow an emergency vehicle to pass \cite{Polizeibericht} (see Table \ref{SECOND}). This creates openings for a further inflow of people. 
\item The overcrowded situation causes dangerous `crowd turbulence' (see Sec. \ref{PaniC}). Many people are falling and pile up on top of each other. Emergency forces cannot reach people quickly enough. 21 of them die of suffocation, and more than 500 are injured \cite{Wiki,asphyxia}.
\item As an unexpectedly large number of people need help, there are not enough emergency forces
at the location of the accident \cite{NotExpected}. Therefore, a triage procedure is applied in the tunnel \cite{Triage}.
As a consequence, many people in critical health conditions do not receive first aid. 
\end{itemize}

\subsection{What Might Have Stopped the Feedback and Cascading Effects}

Overall, one gets the impression that problems occurred on all sides (but we admit that it is easier to identify them afterwards than at the time when decisions must be taken on the basis of often limited and imperfect information). The above analysis shows that things went wrong from the very beginning, and that the situation increasingly got out of control over time. However, we believe that there were also many possibilities to mitigate or overcome problems that contributed to the disaster. Therefore, we will now discuss, how the deadly cascading effect described in the previous subsection might have been stopped or how its size and impact could have been reduced: 
\begin{itemize}
\item One might have been able to find a better suited area for the organization of the event.
\item One could have required higher organizational standards (such as a separation of flow directions). 
\item The decision to hold the event or not could have been taken earlier. This would have facilitated a better preparation and a better coordination. It would also have reduced the commercial and public pressure in case of deciding against the event.
\item Safety and security concerns could have been taken more seriously. The fact that the responsible police officer quit his job \cite{Spiegel} could have been seen as advance warning sign. 
\item Superior contingency plans could have been elaborated, in order to be better prepared for the occurrence of various problems. This applies particularly to the handling of the main bottlenecks of the system: the ramp and the access points.
\item If the evacuation study had raised serious concerns, this might have been able to stop the approval of the event.
\item The various stakeholders could have foreseen larger safety margins and more reserves (also in terms of staff). 
\item It might have been possible to work out a different flow concept, which separates in- and outflows. A circulatory flow organization (where people would come in via the tunnels and both ramps, but leave over the closed-down freeway) would have been interesting to consider. 
\item Obstacles on the ramp (such as the food stand, fences in the way, and police cars) could have been avoided.
\item Efforts could have been made to ensure better communication between the different stakeholders (by reserving [more] priority lines in the mobile phone network) and better communication between the organizers and the crowd (by installing loudspeakers at the ramp and elsewhere). A loudspeaker vehicle could have been moved to the ramp, when it was noticed that no loudspeaker equipment was available on the ramp (around 14:00, see Table \ref{SECOND}), or megaphones could have been used to communicate with the crowd.
\item When it became clear that people had difficulties to enter the festival area and jams formed on the ramp, one might have been able to move the floats further away from the ramp. Moreover, the side ramp could have been used to avoid the jam on the main ramp. 
\item The use of more `pushers' might have been able to increase the outflow from the ramp to the festival area \cite{Spiegel} (but it is not clear how effective this measure would have been, given that the entrance area to the festival ground was quite packed). 
\item More emergency forces (rescue units) could have been positioned on the ramp and next to it.
\item When it was recognized that the crowd management and control did not work as expected, the first police shift might have been extended.
\item When the situation became crowded, cordons could have been established at the isolation devices and at the end of the main ramp.
The outflow of people could have been redirected (either via the side ramp or via the emergency exits). 
\item With loudspeakers or megaphones, people on the overcrowded ramp could have been evacuated earlier and in a more effective way, e.g. by organizing an outflow from the ramp to the festival area behind the chain of police cars that were standing on the ramp (see Fig. \ref{fArea}). Additionally, a continuous evacuation via the staircase could have been established from 16:15 on (or even from 15:31 on, when people needed to use the slopes to get on the festival area) \cite{moveup}. Furthermore, the tunnels could have been used to evacuate the ramp, if the flow directions would have been separated.
\end{itemize}
Given the above alternatives, the crowd disaster might have been avoided in many ways. Already around 13:00 there were first signs that the crowd management concept would not work as planned (see Table \ref{FIRST}. Between 14:30 and 15:15 it was noticed that the ramp constituted a bottleneck that could get out of control. Around 16:25, people climbing the pole, staircase and container were serious warning signs of a critical situation (see Sec. \ref{PaniC}). At this time, it would probably have been possible still to evacuate the ramp, if suitable communication tools had been used. However, the ramp emptied only after 17:00. 

\section{Lessons to be Learned and Recommendations}

\subsection{Summary}

One of the noteworthy points of the Love Parade disaster is that most evidence is available online, which allows many scientists and also the broader public to form an opinion. This dramatically changes the situation compared to many previous disasters, where a lot of evidence is of confidential nature, accessible only to a small number of experts. We believe that the new openness of data can have many beneficial effects on society. This study, for example, hopes to make a contribution to a better understanding of crowd disasters and their avoidance in the future. The accessibility of the materials can also serve organizers of mass events, the police and emergency forces to prepare themselves better.
\par
Through the analysis of publicly available materials and videos, we could identify many factors that have contributed to the Love Parade disaster. Our judgement is that 
the capacity of the area of the mass event already implied various problems, which the organizational concept wanted to overcome by crowd control. However, the delayed start of the event and the unexpected obstruction of the inflow to the festival area from the ramp (i.e. two factors which were probably not anticipated) caused queues that were difficult (or impossible) to manage. Already in the organizational phase, but also in the attempt to manage the flows, many problems came together, and the mutual interaction of these problems made the situation worse. In particular, the cordons that were intended to dissolve the jam at the entrance to the festival area did not yield the expected relief. While they might have worked in case of unidirectional flows, the situation became worse due to the fact that a flow of returning visitors encountered an inflow of arriving people without a separation of the flow directions. From the very beginning, the interaction of many factors resulted in cascading effects, which eventually created a situation that got totally out of control (see Fig. \ref{Cascading}). 
\par
Organizational concepts for mass events are supposed to be robust to the occurrence of single perturbations (`single points of failure'). This in itself, however, does not exclude the possibility that the coincidence or interaction of problems can cause a systemic failure. When certain factors have amplifying effects on other factors (or there are even feedback loops), this can create systemic instabilities. We learn from this that, in order to reach a resilient organization of mass events (and actually any complex system), it is not sufficient to ensure the robustness of each contributing factor. One must also study their {\it interaction} effects, to guarantee that the overall organization is resilient to the coincidence of unfavorable factors as much as possible. 
\par
Our study also sheds new light on issues that have been controversially discussed. Immediately after the Love Parade disaster, the behavior of the crowd and the staircase were blamed for the fatalities. However, our analysis yields a different interpretation: the Love Parade incident shows the typical features of crowd disasters, such as the existence of bottlenecks (and therefore the accumulation of large numbers of people), organizational problems, communication failures, problematic decisions, coordination problems, and the occurrence of crowd turbulence as a result of high crowd densities. 
\par
It is likely that the staircase encouraged a movement of the crowd towards it, when people were trying to escape from the life-threatening density in the ramp area, but  the collective movement seems to have been small (it is not clearly visible in the video recordings). In any case, effective measures (such as an evacuation of the crowd) should have been taken long before critical conditions developed. Given the high density in the ramp area, the occurrence of crowd turbulence or `crowd quakes' was unavoidable. In this dynamical state of the crowd, the lives of people are in serious danger, as people will fall sooner or later. The triggering of this deadly dynamics does not require a particular reason. 
\par
Furthermore, note that the pushing in the crowd at high densities is not necessarily a result of violent behavior, but of the fact that physical forces are transmitted via the bodies of others and adding up. Under such conditions, it is very difficult to keep control over the motion of one's own body, since one is literally moved around by the crowd. The situation in the crowd is difficult also, because no one has an overview of the scene, and the noise level (as well as the overload of the mobile phone network) make communication largely impossible. While the conditions in the crowd were likely to cause a high level of stress, this was a reasonable response to the life-threatening situation. However, a mass panic was most likely {\it not} the cause of the Love Parade disaster. The video recordings from the Love Parade do not provide evidence for a stampede of people, while the dangerous phenomenon of crowd turbulence is clearly visible.
\par
Note that crowd disasters during religious pilgrimage in the past recently led to important insights and also to significant improvements of crowd management and control \cite{crowdsafety1,crowdsafety2,crowdsafety3,crowdsafety4,crowdsafety5}. Many of the lessons learned can also be transferred to other mass events in order to improve their safety. The authors propose to consider the following points (besides the official regulations, of course):
\begin{itemize}
\item Large mass events should preferably take place in locations where experience with the management of large crowds already exists for a long time. It should at least involve some experts who have participated in the organization of previous mass events and know how to handle critical situations. Local organizing teams should be supported by experienced national or supranational professionals.
\item The security concept should be finished, distributed, discussed, and exercised at a pre-specified date well in advance of the event.
\item The event must be planned on the basis of the number of expected people, not on the basis of capacity. 
\item An organizational concept that requires keeping many people out or delays them for hours should be avoided.
\item Facilities (e.g. toilets), supply (particularly food and water), as well as entertainment should be ensured also for people on the way to the festival area and for those waiting to enter.
\item One should implement ways preventing pressure on decisions that may have impact on the safety and security of people. It should not be possible to ignore qualified minority opinions. Contradictory voices should be documented and seriously addressed.
\item Consultants should be encouraged to comment on any critical issues (even beyond the scope of the commissioned analysis). 
\item An analysis of the expected inflows and outflows (and, hence, number of participants) needs to be performed, considering the possibility of large flow variations. A bottleneck analysis is crucial. It must also take into account moving bottlenecks such as floats, but also the operation of police or emergency vehicles.
Confluence, turning and intersection points should be determined. In this context, computer simulations with state-of-the art pedestrian software can be useful, but model parameters must be carefully chosen. Note that computer simulations can often help to identify crowded areas, but they are not sufficient to reveal all kinds of organizational challenges.
\item Critical points should be removed, and it must be checked, whether the remaining problems can be safely handled by crowd management and control measures also under adverse conditions. Safety margins (such as capacity reserves) should be foreseen \cite{crowdsafety1}, and detailed contingency plans should be worked out for likely and unlikely events, and exercised. (Contingency plans serve to reduce the need of improvisation and to ensure a quick and effective response to any occurring problems.) Interaction, cascading and side effects of complicating factors should be analyzed as well. Remaining areas and factors of concern must be continuously monitored (e.g. by video surveillance and special software for real-time analytics \cite{VisualAnalytics}). Sufficient security and emergency forces should always be present to remove or at least mitigate problems early on. Delays in response must be avoided, as they tend to reinforce problems, i.e. quick action is often key to effective counter-measures \cite{QuickDisasterResponse}. To stop possible interaction and cascading effects, suitable decoupling strategies should be implemented.
\item Pressure relief and evacuation strategies must be prepared for any potentially critical areas. Evacuation measures must be started before an area becomes over-crowded. 
\item Intersecting flows should be avoided and different flow directions should be separated (as dense counter-flows are unstable and dangerous \cite{evacuation,counterflows}). A `circular' flow organization, preferably with alternative routes, should be considered \cite{circular}. Moreover, space for emergency vehicles and operations should be reserved. 
\item Fences are not good everywhere. They may turn into obstacles and create dangerous situations. Therefore, the use of fences (or cordons) to stop large numbers of people needs to be carefully considered, as they may be ineffective or deteriorate the situation. In many cases, it is safer to keep people moving (e.g. by re-routing people) rather than stopping them. 
\item Situational awareness and well-functioning communication are crucial. Quick information feedback about the situation in any relevant place and about any relevant factor must be ensured. It is important to have an efficient information flow between the different people and institutions involved (organizers, police, emergency forces, crowd, ...). 
\item In case of problems, the corresponding contingency plan should be applied, and the situation should be continuously (re-)assessed to check for the plausibility of the situational analysis, considering possible alternatives.
\item It should be considered to give police and emergency forces more autonomous (local) decision-making power and responsibility, particularly when communication is interrupted or quick action is needed.
\item Communication must work (both, from a technical and an organizational perspective). It is key to detect, avoid, and respond to critical situations. Communication is also crucial for the capacity to reduce undesirable interaction effects and to stop dangerous cascading effects. 
\item Finally, a safety culture must be actively promoted, reminding everyone that problems can always happen. The motto should be: ``Don't take it easy. Always expect the unexpected!'' Preparations for all sorts of surprising situations (including a sudden change of the weather) should be made as much as possible.
\end{itemize}

\subsection{Some Common Misconceptions}

As discussed before, our study questions a number of common views about crowd disasters. This concerns the following points:
\begin{enumerate}
\item The word `pushing' suggests that people would relentlessly push forward towards their goal, not caring at all about others.
\item The concept of `mass panic' sees a stampede as the origin of the crowd disaster, resulting from a contagious mass psychological effect.
It also assumes that the crowd behaves unreasonably.
\item The term `crushing' suggests that the cause of the crowd disaster is an uncontrolled pushing of a crowd towards a bottleneck, which creates densities so high that the bodies of people are crushed.
\item The word `trampling'  \cite{asphyxia} suggests that people walk carelessly over others.
\end{enumerate}
Such views tend to blame the crowd for the disaster rather than drawing suitable consequences regarding the organization of mass events, the crowd management and communication.
Therefore, recurring disasters may be a consequence of misconceptions about them. 
In contrast to the above interpretations, our analysis of the crowd disaster in Duisburg suggests the following:
\begin{enumerate}
\item {\it It is the `queuing effect' which causes a denser and denser queue of people over time \cite{queuing}, and a lot of pushing in the crowd happens unintentionally.} This is, because physical forces start to add up when the density becomes so high that people start to have body contactŒ. Aggravating factors, which may lead to intentional pushing are (1) long waiting times without food, water, facilities, and entertainment, (2) the absence of understandable, communicated reasons for the delays, and (3) threatening high-density conditions.
\item {\it The main danger are the laws of physics, not psychology  \cite{sim4,turbulence,crowdsafety1,sim2}. People do normally not die because they panic---they panic when their life is in danger.} We do not deny that people get impatient after long waiting times and that some of them also disrespect rules in order to get towards their goal (in particular if these rules do not appear justified to them).  However, even under extremely critical conditions, people helped each other and behaved quite rationally. They overcame barriers, used slopes, staircases, poles and the container mainly, when this was necessary to evacuate themselves and reduce the density in the crowd.  What might have appeared as an unreasonable crowd forcing its way into the festival area may be better interpreted as a crowd trying to find a way out of the dangerous trap it was in. However, despite a rather rational behavior altogether, some individuals suffered from `tunnel vision', which is a phenomenon that can occur under conditions of stress. This becomes evident from the fact that those standing around the poles, staircase and container, hoping to get out, were not considering alternative emergency routes anymore, even when prompted to them by the police \cite{Polizeibericht,wordpress}. 
\item {\it One must distinguish between a `crush' and a `crowd quake', and between (active) trampling and being trampled.} In a classical crush, people are moving towards a physical bottleneck and die in front of its narrowest point. In a `crowd quake', there is typically no systematic flow directions, but people are pushed around by fluctuating forces in the crowd. In Duisburg, people's lives were endangered not by a stampede that crushed other people, but by high crowd pressures (defined as density times variability of body movements \cite{turbulence}). An extreme and fluctuating pressure builds up, when the densities become so high that they cause contact forces between bodies to add up. This ultimately implies the onset of `crowd turbulence'. Under such conditions, the sizes and directions of forces acting on the bodies of visitors move them around in an uncontrolled way, and people have difficulties keeping their balance; when people stumble and fall, this can be the nucleus of a crowd disaster \cite{stumbling} (see next point).
\item {\it When trying to avoid the deadly `domino effect', people may be forced to step on others} \cite{Domino}. In Duisburg, only a few people were relentlessly `crawling' or walking over the heads or shoulders of others. This happened around 16:55, when the ultimate inferno of the crowd disaster happened and it was likely that (some) people had already died. Note, however, that many people probably stepped on others who were lying on the ground. Why did they do such a thing? In a dense and shaky crowd, fallen people have difficulties to get up on their feet again. This may cause a `hole' in the crowd, so that the surrounding people are not anymore counter-balanced: they are pushed from behind, but not anymore from the front. As a consequence, the surrounding people may fall one after another like dominos, causing a pile of people \cite{Domino,FOTO}. If they cannot get back on their feet quickly, they are likely to pass out or suffocate, since they cannot breathe anymore under the weight of others piling up on top of them. Therefore, to avoid falling when pushed around by the crowd, people might be forced to step on others. However, under these conditions, they are rather `walked' than `walking'. That is, while the passive verb ``being trampled'' is correct, the active form ``trampling'' is misleading. 
\end{enumerate}

\subsection{Conclusion and ``Natural Laws'' of Crowd Behavior}

It is obvious that situations such as the ones described above must be absolutely avoided. This requires the choice of a suitable location and an adequate preparation of the mass event, an appropriate organization and crowd management, and a quick response to early warning signs, for which information and communication play a key role. It is also important to understand that crowd behavior follows certain ``laws of nature'', which result from physical, physiological, psychological and social needs of humans such as sufficient space, food, water, and air, toilet facilities, feeling of safety, perceived progress towards the goal, information, communication, entertainment, etc. An insufficient consideration of such factors can promote disasters, particularly if shortcomings accumulate. 

\subsection{Advance Warning Signs of Crowd Disasters} 

To improve the situational awareness of crowd managers, police and emergency forces, Table \ref{Aws} lists a number of successive warning signs of increasingly critical crowd conditions.
\begin{table}[htbp]
{\footnotesize\begin{center}
\begin{tabular}{|l|p{2cm}|p{2.4cm}|p{5.5cm}|}
\hline
 & {\bf Observation} & {\bf Assessment} & {\bf Required Action} \\
\hline
{\bf 0} & Densities are below 2-3 persons per square meter.
& Normal operation at low risk. & Regularly verify normal operation, watch out for perturbations. Make sure that the flow does not exceed the safe value of 82 persons per minute and meter. \\
\hline
{\bf 1} & People accumulate. Certain areas become progressively more crowded. & People slow down due to a bottleneck or stop for some reason. &  Limit inflows to ensure that the expected extent of accumulation will not be exceeded. Gather information and determine the reasons for the accumulation. Prepare possible counter-measures. Move enough security to the respective area. Inform the responsible police and emergency units. \\
\hline
{\bf 2} & Jams of people are forming and growing. & Insufficient outflows may cause serious problems over time (such as high densities), particularly in constraint spaces. & 
Communicate with the crowd. Promptly take appropriate flow reduction measures such as re-directing people. (Keep in mind that stopping people causes a growing pressure in the crowd and impatience.) Move police and emergency units towards the crowded area(s) in case help will be needed. \\
\hline
{\bf 3} & Stop-and-go waves occur (this happens only in dense {\it moving} crowds). People are pushed. & 
The continuous flow has broken down. The outflow capacity is considerably reduced. The situation may escalate quickly. & Take suitable counter-measures. Pressure relief strategies (such as opening emergency routes and re-routing inflows) should be taken and people informed about them. Before, any obstacles (such as fences) in the way must be removed. A sufficient number of emergency units and police must be in the critical area and ready take over control in interaction with the crowd management.  \\
\hline
{\bf 4} & People cannot move freely and are squeezed between others. People are pushed around. & A critical density has built up in the crowd. Injuries can easily happen. & Police should take over control in close consultation with the crowd management. Appropriate contingency plans must be applied. Evacuation is strongly advised. Communication with the crowd is crucial. Emergency forces must be in the most crowded areas, in order to provide first aid whenever needed. \\
\hline
{\bf 5} & People disrespect fences or try to get out of the area. & The situation is critical and likely to get out of control. & Communicate with the crowd and evacuate it. Provide help and first aid. Inform hospitals and additional emergency units about the possibility that the situation may get out of control. \\
\hline
{\bf 6} & Crowd turbulence occurs. People scream or shout for help. & Injuries and fatalities are likely. A crowd disaster can happen any time. & Calm down the crowd and guide it. Continue to evacuate people. Watch out for the areas with the highest densities and largest crowd movements, to ensure support and first aid. Additional emergency vehicles must be called to ensure sufficient manpower, and hospitals must be informed about likely (and potentially many) injuries. \\
\hline
{\bf 7} & People are falling to the ground. People raise arms into the air. & People are in big trouble. Many injuries are to be expected. A crowd disaster is (most likely) happening. & Immediate help and first aid are needed, probably for many people. Hospitals must be prepared to shift from routine to large-scale emergency operation.\\
\hline
{\bf 8} & People crawl over others. & A crowd disaster has probably happened. & Apply rules for a state of serious emergency. \\
\hline
\end{tabular}
\end{center}}
\caption{This table is intended to help assess the level of criticality of the situation in the crowd and take proactive measures to avoid or at least mitigate crowd disasters. Note that at each of these levels, one must already take first preparations for the next one or two (as the situation may change quickly) and communicate the possible scenarios and their implications to all relevant stakeholders. The goal is to de-escalate the situation and get back to lower levels of criticality.}
\label{Aws}
\end{table}

\subsection{Emerging Relevance of Citizen Science and Further Conclusions}

In the subsection above, we have presented science-based suggestions for the avoidance of crowd disasters and an organizational response to critical situations. Deriving these conclusions largely profited from the huge amounts of materials that volunteers have provided, collected, synchronized, and ordered (according to time, locations, content, etc.). This is, where `citizen science' can play an important role. The documentation we have seen from volunteers appears to be more transparent and complete than the information provided by public institutions, and it is better accessible than news from many public and private media (where we often faced the issue that materials could not be retrieved anymore, at least not under the original links). 
\par
Also, scientific institutions would not have had enough resources to do all the documentation work that was performed by these volunteers. However, the collected materials are so
voluminous that one can hardly see the wood for the trees. Therefore, citizen science can largely benefit from an interaction with academic experts. Specialized knowledge is needed to distinguish more relevant from less relevant factors, to interpret empirical evidence, and distinguish more likely from less likely explanations. Besides providing this knowledge, our work also highlights the general and systemic nature of crowd disasters, and it reveals the instabilities (amplification effects) and cascading effects leading to them. 
\par
The systemic nature of many crowd disasters makes their legal handling very difficult, since it is hard to determine the fraction of responsibility that different people and institutions had. However, without a proper response to such systemic failures, people are losing their trust in public institutions, and this undermines their legitimacy \cite{trust}.
\par
Crowd disasters are not the only systemic risk, resulting from interactions and institutional settings that are not suitably designed. The financial crisis is another example \cite{IRGC,queen}, for which nobody seems to be willing to take responsibility. This is mainly, because the individual contributions to it cannot be well quantified. Also human history is full of examples of humanitarian disasters, which happened because nobody felt sufficiently responsible for them. The authors are convinced that the division of responsibility itself is the problem, and that this calls for political and regulatory attention. Scientists could perhaps make a major contribution to the cultural heritage of humanity, if they managed to find new ways to address this fundamental problem \cite{Hilbert}. 

\section*{Acknowledgments}
We would like to thank everyone who has publicly provided materials documenting the Love Parade, and those who have been carefully synchronizing, ordering, analyzing, and describing these materials in hundreds of hours of work. This work has ultimately contributed to the creation of a public good, namely better public safety at future mass events, as it allows many people to learn from  mistakes made in the past. 

\bibliography{loveparade}

\begin{thebibliography}{999}
\item[] Note: Web materials (URL links) were last accessed in February and March 2012. 

\bibitem{control1}
The Scottish Office, 
{\em Guide to Safety at Sports Grounds}
(The Stationery Office, Norwich, 4th edition, 1997).

\bibitem{control2}
Government of Canada, 
{\em Emergency preparedness guidelines for mass, crowd-intensive events},
see \url{http://orise.orau.gov/csepp/documents/planning/evacuation-documents/guidance-documents/canada-crowdevents.pdf} (1994).

\bibitem{control3}
Health and Safety Executive, 
{\em The Event Safety Guide}
(HSE Books, Norwich, 1999).
 
\bibitem{control4}
Health and Safety Executive, 
{\em Managing Crowds Safely}
(HSE Books, Norwich, 2000).

\bibitem{control5}
Health and Safety Executive,
{\em Steps to Risk Assessment. Case Studies}
(HSE Books, Norwich, 1998).

\bibitem{control6}
Independent Street Arts Network, 
{\em Safety Guidance for Street Arts, Carnival, Processions, and Large-Scale Performances}
(Independent Street Arts Network, London, 2004).

\bibitem{control7}
Richtlinie f\"ur Mikroskopische Entfluchtungsanalysen (in German)
The RiMEA Project, see \url{http://www.rimea.de/downloads.html} (2008).

\bibitem{control8}
Musterversammlungsst\"attenverordnung [German Safety Regulations for Public Events], 
see \url{http://www.versammlungsstaettenverordnung.de/vstaettv_neu/}, \url{http://www.is-argebau.de/Dokumente/4231214.pdf}

\bibitem{Predtechenskii}
Predtechenskii, V. M. and Milinskii, A. I.,
{\em Planning for Foot Traffic Flow in Buildings}
(Amerind, New Delhi, (1978).

\bibitem{Fruin} J. J. Fruin, 
The causes and prevention of crowd disasters, 
in  \textit{Engineering for Crowd Safety}, edited by R. A. Smith and J. F. Dickie 
(Elsevier, Amsterdam, 1993), pp. 99-108.

\bibitem{Still}
Still, K.,
Crowd Dynamics, Ph.D. thesis,
University of Warwick, Warwick (2000); see \url{http://www.safercrowds.com/PhD-Chapter-2.html} for the quote given in our manuscript.

\bibitem{evacuation} 
M. Schreckenberg and S. D. Sharma (eds.)  {\em Pedestrian and Evacuation Dynamics}, (Springer-Verlag, Heidelberg, 2002),
see in particular the chapter by D. Helbing, I. Farkas, P. Moln\'{a}r, and T. Vicsek,
Simulation of pedestrian crowds in normal and evacuation situations, pp. 21-58.

\bibitem{TranSci} D. Helbing, L. Buzna, A. Johansson, and T. Werner,
Self-organized pedestrian crowd dynamics: Experiments, simulations, and design solutions, 
\textit{Transportation Science} {\bf 39}(1) (2005) 1--24.

\bibitem{schadenc}
A. Schadschneider, W. Klingsch, H. Kl\"upfel, T. Kretz, C. Rogsch, and A. Seyfried,
Evacuation dynamics: empirical results, modeling and applications,
in: {\it Encyclopedia of Complexity and Systems Science} {\bf 3}, pp. 3142ff (Springer, Berlin, 2009).

\bibitem{encyklopedia} D. Helbing and A. Johansson, Pedestrian, Crowd and Evacuation Dynamics, in: {\it Encyclopedia of Complexity and Systems Science} {\bf 16}, 6476-6495 (2010).

\bibitem{crowdsafety1} A. Johansson, D. Helbing, H. Z. A-Abideen, and S. Al-Bosta (2008) From crowd dynamics to crowd safety: A video-based analysis. {\it Advances in Complex Systems} {\it 11}(4), 497-527,
see also \url{http://www.trafficforum.org/crowdturbulence}.
\bibitem{crowdsafety2}
K. Haase, Scheduling of Hajjis Groups in Hajj 1427H, Specialized Architectural, Engineering \& Technical Reviewed Magazine issued by Ministry of Municipal \& Rural Affairs, No. 10, pp. 70-75, Saudi Arabia (2006).
\bibitem{crowdsafety3}
K. Haase, Scheduling and re-scheduling the departure and stoning times of the Hajjis in 1428H, Specialized Architectural, Engineering \&Technical Reviewed Magazine issued by Ministry of Municipal \& Rural Affairs, Saudi Arabia (2007).
\bibitem{crowdsafety4}
D. Serwill, R. Vollmer, S. Al Bosta, A.Tayara: Design and Organization of the Jamarat Stoning Process at Hajj 1427-1430H, IVV (Ingenieurgruppe IVV), Aachen-Riyadh (10/2009).
\bibitem{crowdsafety5} H. H. Al Nabulsi, {\it How Can Vulnerability and Risk Be Reduced in Large-Scale Gatherings? An Assessment of Vulnerability and Risk in Mass Gatherings}  (MA thesis, School of Built Environment, Oxford Brookes University, Oxford, September 2009).

\bibitem{exp1}
D. Helbing, M. Isobe, T. Nagatani, and K. Takimoto, Lattice gas simulation of experimentally studied evacuation dynamics, {\it Physical Review E} {\bf 67}, 067101 (2003);
M. Isobe, D. Helbing, and T. Nagatani, Experiment, theory, and simulation of the evacuation of a room without visibility,  {\em Physical Review E} {\bf 69}, 066132 (2004).

\bibitem{exp2}
A. Kirchner {\it et al.}, Simulation of competitive egress behavior: comparison with aircraft evacuation data, {\it Physica A} {\bf 324}, 689-697 (2003).

\bibitem{exp3} S.P. Hoogendoorn and W. Daamen, Pedestrian behavior at bottlenecks, {\it Transportation Science} {\bf 39}(2), 147-159 (2005).


\bibitem{exp4} D. Nilsson and A. Johansson,  Social influence during the initial phase of a fire evacuation---Analysis of evacuation experiments in a cinema theatre, {\it Fire Safety Journal} {\bf 44}(1), 71-79 (2009).  

\bibitem{exp5} A. Seyfried {\it et al.}  New insights into pedestrian flow through bottlenecks, {\it Transportation Science} {\bf 43}(3), 395-406 (2009). 

\bibitem{exp6} W. Klingsch, C. Rogsch, A. Schadschneider, and M. Schreckenberg (eds.)
{\it Pedestrian and Evacuation Dynamics 2008} (Springer, Berlin, 2010)

\bibitem{exp7} Z. Fang, W. Song, J. Zhang, H. Wu, Experiment and modeling of exit-selecting behaviors during a building evacuation, {\it Physica A} {\bf 389},  815-824 (2010).

\bibitem{exp8} W. Daamen and S.P. Hoogendoorn,  Emergency door capacity: influence of door width, population composition and stress level, {\it Fire Technology} {\bf 48}(1), 55-71 (2012).
 
\bibitem{sim1}
D. Helbing, A mathematical model for the behavior of pedestrians. {\it Behavioral Science} {\bf 36}, 298-310 (1991).

\bibitem{sim2}
M. Moussa\"{\i}d, D. Helbing, and G. Theraulaz, How simple rules determine pedestrian behavior and crowd disasters. {\it Proceedings of the National Academy of Sciences of the USA (PNAS)} {\bf 108}(17), 6884-6888 (2011).

\bibitem{sim3}
D. Helbing and P. Moln\'{a}r, 
Social force model for pedestrian dynamics,
{\it Physical Review E} {\bf 51} 4282-4286 (1995).

\bibitem{sim4}
D. Helbing, I. Farkas, I. and T. Vicsek,
Simulating dynamical features of escape panic,
{\it Nature} {\bf 407} 487-490 (2000).

\bibitem{sim5}
C. Burstedde, K. Klauck, A. Schadschneider, and J. Zittartz, 
Simulation of pedestrian dynamics using a two-dimensional cellular automaton, 
{\em Physica A} {\bf 295}(4), 507-525 (2001).

\bibitem{sim6} T. Nagatani, The physics of traffic jams, {\it Reports on Progress in Physics} {\bf 65}, 1331-1385 (2002).

\bibitem{sim7} A. Kirchner and A. Schadschneider,  Simulation of evacuation processes using a bionics-inspired cellular automaton model for pedestrian dynamics, {\it Physica A} {\bf 312}, 260-276 (2002).

\bibitem{sim8} S. Hoogendoorn,  Pedestrian flow modeling by adaptive control. {\it Transportation Research Record} {\bf 1878}, 95-103 (2004).

\bibitem{sim9}
D. Helbing, A. Johansson, J. Mathiesen, M. H. Jensen, and A. Hansen, 
Analytical approach to continuous and intermittent bottleneck flows,
{\it Physical Review Letters} {\bf 97}, 168001 (2006).

\bibitem{sim10} G. Antonini, M. Bierlaire, and M. Weber, Discrete choice models of pedestrian walking behavior. {\it Transportation Research B} {\bf 40}(8):667-687 (2006).
 
\bibitem{sim11}
W. Yu and A. Johansson, 
Modeling crowd turbulence by many-particle simulations,
{\it Phys. Rev. E} {\bf 76}, 046105 (2007). 

\bibitem{sim12} M. Moussa\"{\i}d {\it et al.}, Experimental study of the behavioural mechanisms underlying self-organization in human crowds. {\it Proceedings of the Royal Society B} {\bf 276}, 2755-2762 (2009).

\bibitem{Polizeibericht} Polizeipr\"asidium Essen, Vorl\"aufiger Abschlussbericht from 31. Oktober  2010 zur Nachbereitung des polizeilichen Einsatzes der Veranstaltung ``Loveparade'' am 24.07.2010 in Duisburg [Police Report], see \url{http://www.mik.nrw.de/fileadmin/user_upload/Redakteure/Dokumente/Themen_und_Aufgaben/Schutz_und_Sicherheit/110601_vorlaeufiger-abschlussbericht.pdf}

\bibitem{FinalReport} Final report of the city of Duisburg, see \url{http://www.duisburg.de/ratsinformationssystem/bi/vo0050.php?__kvonr=20056110&voselect=20049862}

\bibitem{Earth} For a Google Earth photograph of the Love Parade festival area in Duisburg and its surrounding, see \url{http://www.earth-dots.de/massenpanik-auf-der-love-parade-99843.html}. For an aerial photograph taken on the day of the event see \url{http://www1.wdr.de/themen/archiv/sp_loveparade/loveparade112.html}

\bibitem{360} 360 degree representation of the site of the event: \url{http://www1.wdr.de/themen/archiv/sp_loveparade/loveparade150.html};  a Google Streetmap version can be found here: \url{http://g.co/maps/65r7j} 

\bibitem{youtube} YouTube channel with videos of the Love Parade in Duisburg, see \url{http://www.youtube.com/user/LoveparadeDuisburg?feature=watch} 

\bibitem{Wiki} Wikipedia article on the Love Parade Disaster: \url{http://en.wikipedia.org/wiki/Love_Parade_stampede}

\bibitem{Wikipedia} Wikipedia article on the Love Parade and its history: \url{http://en.wikipedia.org/wiki/Love_Parade}

\bibitem{wikikarte} Wikipedia map associated with German Wikipedia article: \url{http://en.wikipedia.org/wiki/File:%C3%9Cbersichtskarte_Loveparade_Duisburg_2010.jpg}

\bibitem{Leaks} Loveparade 2010 Duisburg planning documents, 2007-2010, see \url{http://wikileaks.org/wiki/Loveparade_2010_Duisburg_planning_documents,_2007-2010}

\bibitem{responsible} Expert opinion on the areas of responsibility of the organizer, city of Duisburg, and police at the Love Parade 2010, see 
\url{http://www.mik.nrw.de/fileadmin/user_upload/Redakteure/Dokumente/Themen_und_Aufgaben/Schutz_und_Sicherheit/100901gutachten_sfe.pdf}

\bibitem{Dok} \url{http://loveparade2010doku.wordpress.com/links/}

\bibitem{Einsatz} Event Log of the City of Duisburg, \url{http://loveparade2010doku.files.wordpress.com/2010/10/10-1405_2_anlage_621.pdf}

\bibitem{Entfluchtung} 
Entfluchtungsanalyse Love Parade 2010 [Evacuation analysis Love Parade 2010], 
see \url{http://docunews.org/go/2010_07_13-traffgo-ht-entfluchtungsanalyse-loveparade-2010-49/},
Entfluchtungsanalyse Loveparade 2010, Nachtrag [Supplement], 
see Anlage 30 [attachment no. 30] at \url{https://www.duisburg.de/ratsinformationssystem/bi/vo0050.php?__kvonr=20056110&voselect=20049862}

\bibitem{wordpress} Documentation of the events during the Love Parade 2010 in Duisburg, see \url{http://loveparade2010doku.wordpress.com}

\bibitem{Videos} Synchronized video collection, with information regarding location and content, see \url{http://loveparade2010doku.wordpress.com/2010/07/31/loveparade-ungluck-videos-verknupfen-zeitstrahl/} . The synchronization procedure is described in \url{http://loveparade2010doku.files.wordpress.com/2012/02/lopa_zeitstrahl-verknuepfungen_v120218.pdf}.

\bibitem{Spiegel} A. Brandt, G. B\"onisch, J. Dahlkamp, and S. R\"obel, Schwarzer Samstag, Der Spiegel 20/2011, pp. 59-69, see \url{http://www.spiegel.de/spiegel/print/d-78522270.html}

\bibitem{Tagesschau} Tagesschau Love Parade Dossier, see \url{http://www.tagesschau.de/inland/dossierloveparade100.html}

\bibitem{WDRmediathek} WDR Love Parade Dossier, see \url{http://www.wdr.de/mediathek/html/regional/uebersicht/katastrophe-loveparade.xml}

\bibitem{SpiegelTV} Spiegel TV documentation on the Love Parade, see \url{http://www.spiegel.de/video/video-1077726.html}

\bibitem{SpiegelVideoDoku}
Spiegel TV documentation on the Love Parade: \\
\url{http://www.youtube.com/watch?v=Opd0rZVsspQ} \\ 
\url{http://www.youtube.com/watch?v=DTbJ_vbT8Cw} \\
\url{http://www.youtube.com/watch?v=S9ILNAv0J1A} \\
\url{http://www.youtube.com/watch?v=FHD8aqsCr9U}

\bibitem{RTL}
RTL II video documentation ``100 Tage'' [100 days later] on the Love Parade: \\
1/15 \url{http://www.youtube.com/watch?v=fy1NDX_nA3M} \\
2/15 \url{http://www.youtube.com/watch?v=38iqVvS_Ljs} \\
3/15 \url{http://www.youtube.com/watch?v=6zhX6bX2y7s} \\
4/15 \url{http://www.youtube.com/watch?v=G_UJ8c97kxg} \\
5/15 \url{http://www.youtube.com/watch?v=2ZVzlrhXeBw} \\
6/15 \url{http://www.youtube.com/watch?v=T5qNYmpD5mk} \\
7/15 \url{http://www.youtube.com/watch?v=KKqXMvAWprc} \\
8/15 \url{http://www.youtube.com/watch?v=SgI2aKofkLk} \\
9/15 \url{http://www.youtube.com/watch?v=0y00qe3PPsc} \\
10/15 \url{http://www.youtube.com/watch?v=4azc7lPhVnU} \\
11/15 \url{http://www.youtube.com/watch?v=HRC7S4qlvjs} \\
12/15 \url{http://www.youtube.com/watch?v=cnr7S22_BM8} \\
13/15 \url{http://www.youtube.com/watch?v=_RZqYhzekjs} \\
14/15 \url{http://www.youtube.com/watch?v=AusbCbphWAo} \\
15/15 \url{http://www.youtube.com/watch?v=mVYTlhkcKDY}

\bibitem{LopaventVideos} 
Video documentation of the organizers:\\
\url{http://www.youtube.com/watch?v=8y73-7lFBNE} (in English)\\
\url{http://live.loveparade.com/fkxt76kdrf887t/videos/dokumentarfilm_hires.mp4} (in German)


\bibitem{overv} Synchronized multi-perspective video, see \url{http://www.youtube.com/watch?v=up95bUU3L0M}

\bibitem{syncro} Overview videos cut from many videos of visitors, see: \\
\url{http://www.youtube.com/watch?v=V9cbqu5sEE0}\\ 
\url{http://www.youtube.com/watch?v=cDJaAvF0l7s}\\ 
\url{http://www.youtube.com/watch?v=sBE79UoxCF4}\\ 
\url{http://www.youtube.com/watch?v=qJscpcZC45s}\\
\url{http://www.youtube.com/watch?v=vooMCrcOXGs}\\
\url{http://www.youtube.com/watch?v=pOx_VHJd6G4} 

\bibitem{VideoDoku} Video documentation by `LoveparadeDuisburg': \\
Part 1: \url{http://www.youtube.com/watch?v=BRUlHnvJl-Q} \\ 
Part 2: \url{http://www.youtube.com/watch?v=VDOlXcobbJM} \\
Part 3: \url{http://www.youtube.com/watch?v=y_agoPlP_dA} \\ 
Part 4: \url{http://www.youtube.com/watch?v=pfxdSZg9KWk} 


\bibitem{Critique} Critical summary videos,
see \url{http://www.youtube.com/watch?v=NxGSrPALytM}
and \url{http://www.youtube.com/watch?v=up95bUU3L0M} 


\bibitem{Satellite} Video illustrating the events with comments on a satellite image, see \url{http://www.youtube.com/watch?v=TC0Ygi7AMfM}

\bibitem{Surveillance} Incomplete collection of surveillance videos of the organizer of the Love Parade, see \url{http://loveparade2010doku.wordpress.com/2010/08/30/lopavent-veroffentlicht-originalvideos-von-7-der-16-uberwachungskameras-der-loveparade-2010/#downloads}. Camera 14 and 15 covered the tunnel in the West, camera 16 the one in the East. Camera 13 provided a view over the ramp. Camera 12 shows the upper part of the ramp and the part where it enters the festival area. Camera 4 shows the view from the stage on the festival area, sometimes pointing approximately towards the ramp. The floats and the surrounding people are shown by cameras 4,5 and 12. An overview of 6 surveillance cameras it given at:\\ 
\url{http://live.loveparade.com/fkxt76kdrf887t/videos/kameras/6ersplit/hires/6erSplit_HiRes_1530_1540.mp4}\\
\url{http://live.loveparade.com/fkxt76kdrf887t/videos/kameras/6ersplit/hires/6erSplit_HiRes_1540_1600.mp4}\\
\url{http://live.loveparade.com/fkxt76kdrf887t/videos/kameras/6ersplit/hires/6erSplit_HiRes_1600_1620.mp4}\\
\url{http://live.loveparade.com/fkxt76kdrf887t/videos/kameras/6ersplit/hires/6erSplit_HiRes_1620_1640.mp4}

\bibitem{timeline} Time line: \url{http://loveparade2010doku.wordpress.com/2010/07/28/loveparade-2010-zeitablauf-sperrungen-und-durchlassungen/},
\url{http://blog.odem.org/2010/07/ablauf-tragoedie.html}

\bibitem{videocoll} Collection of videos, time-ordered, with approximate geo-location of filming person: \url{http://loveparadevideos.heroku.com/}


\bibitem{Zeitstrahl}
Before 16:07:30 CURUBA300 \url{http://www.youtube.com/watch?v=Eaq2DPmafyI} \\
16:15:52 Todesparade2010 1/10 \url{http://www.youtube.com/watch?v=kVpkclRCXaQ} \\
16:16:35 trampelklette 1 \url{http://www.youtube.com/watch?v=KaDoWMAZYyo} \\
16:22:54 xetzxetzxetz 16:25 \url{http://www.youtube.com/watch?v=P3jL17JOA8U} \\
16:23:45 thaido \url{http://www.youtube.com/watch?v=6SXgp3VlM88}\\
16:24:12 Todesparade2010 3/10 \url{http://www.youtube.com/watch?v=_PQqBePT6ig}\\
16:24:38 GermanTranceMusic 1 \url{http://www.youtube.com/watch?v=yNcIlNGQkgk}\\
16:24:48 pizzamanne 2 [from 16:24] \url{http://www.youtube.com/watch?v=8aQuTaMbS38}\\
16:24:54 trampelklette 2 \url{http://www.youtube.com/watch?v=MSGJW1uJMS0}\\
16:25:29 xetzxetzxetz 16:27 \url{http://www.youtube.com/watch?v=MvlzywaFnmc}\\
16:26:07 Todesparade2010 4/10 \url{http://www.youtube.com/watch?v=so6-7Ezeo3U}\\
16:26:20 Letroen \url{http://www.youtube.com/watch?v=PMld0jhO7Jk}\\
16:28:24 pizzamanne \url{http://www.youtube.com/watch?v=8aQuTaMbS38}\\
16:29:01 dirkbvb74-1 \url{http://www.youtube.com/watch?v=YmQR6kgwSxA}\\
16:29:12 coolwojtek 1 \url{http://www.youtube.com/watch?v=wsOyIBCMExM}\\
16:29:40 funka84 \url{http://www.youtube.com/watch?v=t3nDQti-zDY}\\
16:29:44 Todesparade2010 6/10 \url{http://www.youtube.com/watch?v=WTiQ131QejE}\\
16:30:33 pizzamanne 2 [from 16:30] (starting at 2:15) \url{http://www.youtube.com/watch?v=8aQuTaMbS38}\\
16:31:21 TheHubschrauber92 \url{http://www.youtube.com/watch?v=cPkemW3ixC4}\\
16:31:59 rheinhousen 2 \url{http://www.youtube.com/watch?v=UBkilEAyK80}\\
16:32:10 dreamshockerPro \url{http://www.youtube.com/watch?v=adZ6RaK8C-E}\\
16:33:11 Habanos2009 \url{http://www.youtube.com/watch?v=EoX1rPefuuI}\\
16:33:55 joelooo1 1 \url{http://www.youtube.com/watch?v=rVvUl1LVEwQ}\\
16:34:49 joelooo1 2 \url{http://www.youtube.com/watch?v=Mw4W6z_wWcM}\\
16:34:51 xfugox \url{http://www.youtube.com/watch?v=arncFs1DU0k}\\
16:35:11 pizzamanne 2 [from 16:35] \url{http://www.youtube.com/watch?v=8aQuTaMbS38}\\
16:35:19 Deepkisses [from 16:23] \url{http://www.youtube.com/watch?v=3XFTVWUN8nw}\\
16:35:32 dirkbvb74-2 \url{http://www.youtube.com/watch?v=r1toPUusRGU}\\
16:36:09 MarkusLedwig \url{http://www.youtube.com/watch?v=h3ik6n2BPa8}\\
16:37:06 Ko0rn \url{http://www.youtube.com/watch?v=m597wnEMUTo}\\
16:37:11 xetxetzxetz 16:38 \url{http://www.youtube.com/watch?v=mo97xHkHh3w}\\
16:37:36 Cenobite1988 \url{http://www.youtube.com/watch?v=UqV3Nx0MFh4}\\
16:37:44 real02 3/3 \url{http://www.myvideo.de/watch/7673480/Massenpanik_Loveparade2010_3_3}\\
16:37:57 GSayMusic \url{http://www.youtube.com/watch?v=kUEB4OcsoRM} (removed)\\
16:37:58 Klabauter0815 \url{http://www.youtube.com/watch?v=B1usiOJRWDU} (removed)\\
16:38:16 Goonies11000 \url{http://www.youtube.com/watch?v=6dReLGi1lnc}\\
16:38:23 pizzamanne 2 [from 16:38] \url{http://www.youtube.com/watch?v=8aQuTaMbS38}\\
16:38:28 aggrostar69 \url{http://www.youtube.com/watch?v=-2onoJLq2-8}\\
16:38:50 dirkbvb74-3 \url{http://www.youtube.com/watch?v=UWXXDEZ4oKg}\\
16:39:01 Deepkisses 2 \url{http://www.youtube.com/watch?v=3XFTVWUN8nw}\\
16:39:20 SpaceCommander77 \url{http://www.youtube.com/watch?v=gLdhwP4_IXM}\\
16:39:26 FrEaKyLaDiiiEs \url{http://www.youtube.com/watch?v=d1la2QTDBPU}\\
16:39:38 GoofyMcPott \url{http://www.youtube.com/watch?v=OOjNklbvxA8}\\
16:41:27 pizzamanne 3 [from 16:41] \url{http://www.youtube.com/watch?v=9DvH1BYFVCQ}\\
16:41:40 SKAKnabe \url{http://www.youtube.com/watch?v=prXmQBf27WI} (removed)


\bibitem{pizzamanne}
Collection of synchronized videos by `pizzamanne' (potentially incorrect time stamp on videos in brackets): \\
        1017 - ca. 15:38 (15:38 / 15:38:38) -
        \url{http://www.youtube.com/watch?v=OGd1N6_hm1k} \\
        1019 - ca. 15:58 (15:58 / 15:58) -
        \url{http://www.youtube.com/watch?v=vdJzV26tL78} \\
        1022 - ca. 16:07 (16:07 / 16:07:12) -
        \url{http://www.youtube.com/watch?v=2_LfccdiSN0} \\
        1024 - ca. 16:08 (16:08; 16:08:49) -
        \url{http://www.youtube.com/watch?v=-4f8AMEstkM} \\
        1029 - ca. 16:11 (16:11 / 16:11:53) -
        \url{http://www.youtube.com/watch?v=_Zsmr9CelDY} \\
        1033 - ca. 16:15 (16:15 / 16:15:36) -
        \url{http://www.youtube.com/watch?v=_gJP2QVbwVs} \\
        1034 - ca. 16:17 (16:17 / 16:17) -
        \url{http://www.youtube.com/watch?v=pweaCIXTs5U} \\
        1035 - ca. 16:18 (16:18 / 16:18) -
        \url{http://www.youtube.com/watch?v=f00Mx8MY6y8} \\
        1037 - 16:22:55 (16:22 / 16:22:54) -
        \url{http://www.youtube.com/watch?v=Lkc7zgPv8y0} \\
        1038 - 16:24:49 (16:24 / 16:24:48) -
        \url{http://www.youtube.com/watch?v=9wpVvSWa4_8} \\
        1039 - 16:28:24 (16:28 / 16:28:24) -
        \url{http://www.youtube.com/watch?v=NmEcLa4OfvU} \\
        1040 - 16:30:33 (16:30 / 16:30:34) -
        \url{http://www.youtube.com/watch?v=Ra_j4kTOqqE} \\
        1041 - 16:35:11 (16:35 / 16:35:12) -
        \url{http://www.youtube.com/watch?v=AYjlXhPcF4s} \\
        1043 - ca. 16:37 (16:37 / 16:37) - \\ 
        \url{http://www.youtube.com/watch?v=Od-DGsPlues} \\
        1044 - 16:38:22 (? / 16:38:24) -
        \url{http://www.youtube.com/watch?v=Y3qlzW-B9OM} \\
        1054 - 16:59:23 (16:59 / 16:59:35) -
        \url{http://www.youtube.com/watch?v=Uk9BXDdGJoQ} \\
        1055 - ca. 17:04:43 (?) (17:04 / 17:04:43) -
        \url{http://www.youtube.com/watch?v=qUBjVtjwl3o} \\
        1056 - ca. 17:06 (17:06 / 17:06) - \\ Schnipsel von 1 Sek. -
        \url{http://www.youtube.com/watch?v=oCyzQ4NgVBU} \\
        1057 - 17:08:40 (17:08 / 17:08:50) -
        \url{http://www.youtube.com/watch?v=iJ4NdwXa0Os} \\
        1058 - ca. 17:23 (17:23 / 17:23) -
        \url{http://www.youtube.com/watch?v=qtyC1YvwbEc} \\
        1065 - ca. 20:38 (20:38 / 20:38:35) -
        \url{http://www.youtube.com/watch?v=UEFnUi5w7gs} \\
        1067 - ca. 20:47 (20:47 / 20:46:55) -
        \url{http://www.youtube.com/watch?v=ZsUHVV_k3r0} \\
        1068 - ca. 20:49 (20:49 / 20:49:31) -
        \url{http://www.youtube.com/watch?v=3UYacod0w6g}


\bibitem{Pizzamanne}
Collection of synchronized multi-view videos, by `pizzamanne': \\
Part 1 (15:20-16:25): \url{http://www.youtube.com/watch?v=V9cbqu5sEE0} \\
Part 2 (16:25-16:34): \url{http://www.youtube.com/watch?v=cDJaAvF0l7s} \\
Part 3 (16:34-16:44): \url{http://www.youtube.com/watch?v=sBE79UoxCF4} \\
Part 4 (16:44-16:54): \url{http://www.youtube.com/watch?v=qJscpcZC45s} \\
Part 5 (16:54-17:05): \url{http://www.youtube.com/watch?v=vooMCrcOXGs} \\
Part 6 (17:05-17:20): \url{http://www.youtube.com/watch?v=pOx_VHJd6G4}

\bibitem{kaydee271}
Video collection of `kaydee271': \\
    1/8 - ca. 15:25 -
    \url{http://www.youtube.com/watch?v=2fT7qKC8QOw} \\
    2/8 - ca. 15:35 -
    \url{http://www.youtube.com/watch?v=rrpYKGNlhdw} \\
    3/8 - 17:03:34 -
    \url{http://www.youtube.com/watch?v=IoxPIvrFCNg} \\
    4/8 - 17:05:14 -
    \url{http://www.youtube.com/watch?v=t3_3UIZS3dw} \\
    5/8 - 17:09:41 -
    \url{http://www.youtube.com/watch?v=Kc8wEMiOxoo} \\
    6/8 - 17:11:19 -
    \url{http://www.youtube.com/watch?v=gN4NNmxtRU4} \\
    7/8 - 17:12:08 -
    \url{http://www.youtube.com/watch?v=JfaLr_Y4U18} \\
    8/8 - 17:15:15 -
    \url{http://www.youtube.com/watch?v=FYXQLgd_VA8}

\bibitem{backtony}
Video collection of `tr1nd' ('Backtony'): \\
\url{http://www.youtube.com/watch?v=vAxOeGUCZNI}

\bibitem{real02}
Video collection of `real02': \\
        3/3 - 16:37:44 - \url{http://www.youtube.com/watch?v=IdtR4Ks_n6I}, also at
        \url{http://www.myvideo.de/watch/7673480/Massenpanik_Loveparade2010_3_3} \\
        1/3 - 16:58:19 - \url{http://www.youtube.com/watch?v=dP-5VOCU30U}, also at
        \url{http://www.myvideo.de/watch/7673149/Massenpanik_Loveparade2010_1_3},\\
        2/3 - 17:05:52 - \url{http://www.youtube.com/watch?v=HvXwrbCjgf4}, also at
        \url{http://www.myvideo.de/watch/7673325/Massenpanik_Loveparade2010_2_3}


\bibitem{hitower}
Video collection of `hitower78':  \\
        1 - 16:36:17 (16:08:43 / 16:36:21 [16:37:27]) -
        \url{http://www.youtube.com/watch?v=0ioEPdfvZdw} \\
        2 - 16:50:23 (16:22:49 / 16:50:27) -
        \url{http://www.youtube.com/watch?v=u2r1AFSuHNI} \\
        3 - 16:52:33 (16:24:59 / 16:52:37) -
        \url{http://www.youtube.com/watch?v=IEtFZVOjsK4} 


\bibitem{Todesparade}
Video collection of `Todesparade2010': \\
        1/10 - 16:15:52 (16:16)
        \url{http://www.youtube.com/watch?v=kVpkclRCXaQ} \\
        2/10 - ca. 16:22 (16:23)
        \url{http://www.youtube.com/watch?v=lFxyl9OaqHk} \\
        3/10 - 16:24:12 (16:25)
        \url{http://www.youtube.com/watch?v=_PQqBePT6ig} \\
        4/10 - 16:26:07 (16:27)
        \url{http://www.youtube.com/watch?v=so6-7Ezeo3U} \\
        5/10 - ca. 16:29 (16:30)
        \url{http://www.youtube.com/watch?v=VohMiM54wpA} \\
        6/10 - 16:29:44 (16:31) 
        \url{http://www.youtube.com/watch?v=WTiQ131QejE} \\
        7/10 - ca. 16:31 (16:32)
        \url{http://www.youtube.com/watch?v=qnZGcnWyJOM} \\
        8/10 - ca. 16:33 (16:34)
        \url{http://www.youtube.com/watch?v=3cVyb3W-GAM} \\
        9/10 - ca. 16:52 (16:53) 
        \url{http://www.youtube.com/watch?v=hPZiHkZQH98} \\
        10/10 - ca. 16:53 (16:54)
        \url{http://www.youtube.com/watch?v=2Pl1dVP8bA4}

\bibitem{goonies}
Video collection of  `goonies11000': \\
        1 - 16:38:15 -
        \url{http://www.youtube.com/watch?v=6dReLGi1lnc} \\
        2 - 16:55:51 -
        \url{http://www.youtube.com/watch?v=L9wfsupvD24} \\
        3 - 16:59:41 -
        \url{http://www.youtube.com/watch?v=rzz5geLWPV4} \\
        4 - 17:18:35 -
        \url{http://www.youtube.com/watch?v=qbQr5yAjqxs}

\bibitem{coolwojtek}
Video collection of  `coolwojtek': \\
    1 - 16:29:12 (16:28:42) -
    \url{http://www.youtube.com/watch?v=wsOyIBCMExM} \\
    2 - 16:44:44 (16:44:14) -
    \url{http://www.youtube.com/watch?v=8aTXT3ht8VU} \\
    3 - 16:48:34 (16:48:04) -
   \url{http://www.youtube.com/watch?v=1lHLIn78650} \\
    4 - 16:51:02 (16:50:32) -
    \url{http://www.youtube.com/watch?v=XKo3rb5R_Dc} \\
    5 - 16:54:36 (15:53:36) -
    \url{http://www.youtube.com/watch?v=xxd_KlaCiNY}

\bibitem{mbreezer}
Video collection of `mbreezer': \\
Summary - 
    \url{http://www.youtube.com/watch?v=OfQjXi3J3ns} \\
    0 - 16:38:36 -
    \url{http://www.youtube.com/watch?v=5uio5rSv520} \\
    1 - 16:47:31 -
    \url{http://www.youtube.com/watch?v=rgvG2Pp7yfc}  \\
    2 - 16:51:41 -
    \url{http://www.youtube.com/watch?v=ofFd928JY6k} \\
    3 - 16:53:07 -
    \url{http://www.youtube.com/watch?v=3aodLRaPu0E} \\
    4 - 16:53:50 -
    \url{http://www.youtube.com/watch?v=u8N3ZcqOgfU} \\
    5 - 16:55:05 -
    \url{http://www.youtube.com/watch?v=JdblWBmaM4Q} \\
    6 - 17:00:30 -
    \url{http://www.youtube.com/watch?v=VTF0v-WY934} \\
    7 - 17:02:47 -
    \url{http://www.youtube.com/watch?v=V-WKhjRMcRM} \\
    8 - 17:03:11 -
    \url{http://www.youtube.com/watch?v=s2ywRwxfHbo} \\
    9 - 17:05:48 -
    \url{http://www.youtube.com/watch?v=52KPyHrqkwY}

\bibitem{hell}
Video collection of `The1art1of1hell':\\
    1 - 16:59:05 (16-58-09) -
    \url{http://www.youtube.com/watch?v=YxIZwvvhpp8} \\
    2 -  17:03:28 (17-02-30) -
    \url{http://www.youtube.com/watch?v=BbCrmUZeJoY} \\
    3 - 17:05:38 (17-03-33) -
    \url{http://www.youtube.com/watch?v=NLC3vyp0b9U} \\
    4 - 17:07:30 (17-05-42) -
    \url{http://www.youtube.com/watch?v=QcAVAYDolKc} \\
    5 - 17:08:48 (17-07-34) -
    \url{http://www.youtube.com/watch?v=-iCA6g984aY} \\
    6 - 17:09:52 (17-08-51) -
    \url{http://www.youtube.com/watch?v=LXLuoqNDJ-A} \\
    7 - 17:11:26 (17-09-31) -
    \url{http://www.youtube.com/watch?v=-7WDbM4EonY} \\
    8 - 17:12:28 (17-11-30) -
    \url{http://www.youtube.com/watch?v=zCRGtjcbyw8} \\
    9 - 17:13:47 (17-12-32) -
    \url{http://www.youtube.com/watch?v=gEa9_k-Th7U} \\
    10 - 17:14:50 (17-13-51) -
    \url{http://www.youtube.com/watch?v=_CiEkTPk9BQ}

\bibitem{rkjorge70}
Video collection of `rkjorge70':\\
    1 - 17:10:44  -
    \url{http://www.youtube.com/watch?v=Nb0m_n0KGms} \\
    2 - 17:13:26 -
    \url{http://www.youtube.com/watch?v=qQbto4MNqOU} \\
    3 - 17:19:41 -
    \url{http://www.youtube.com/watch?v=WR56iEfBQoo}

\bibitem{Berlin} Pictures of the Love Parade in Berlin, see:
 \url{http://www.skyscrapercity.com/showpost.php?s=dabeccde05e177be86cfa1009c0e245e&p=60946437&postcount=44}

\bibitem{Pope} See \url{http://de.wikipedia.org/wiki/Weltjugendtag_2005},\\
\url{http://en.wikipedia.org/wiki/Marienfeld},\\
\url{http://commons.wikimedia.org/wiki/Category:Marienfeld_%28World_Youth_Day%29?uselang=de}

\bibitem{political} The event was politically desired, see \url{http://www.duisboard.de/forum/index.php?page=Thread&postID=113125} 

\bibitem{Weidmann}
Weidmann, U.,
{\it Transporttechnik der Fu{\ss}g\"anger} (Schriftenreihe des Institut f\"ur Verkehrsplanung, Transporttechnik, Stra{\ss}en- und Eisenbahnbau {\bf 90}, ETH Z\"urich, Z\"urich, 1993).

\bibitem{HCM} Transportation Research Board, {\it Highway Capacity
Manual}, Special Report 209 (Transportation Research Board, Washington DC, 1985), see the chapter on pedestrians.

\bibitem{LOS1}
Polus, A., Schofer, J. L., Ushpiz, A.,
Pedestrian flow and level of service,
{\it Journal of Transportation Engineering} {\bf 109} (1983) 46--56.

\bibitem{LOS2}
Fruin, J. J.,
Designing for pedestrians: A level-of-service concept,
\textit{Highway Research Record} {\bf 355} (1971) 1--15. 

\bibitem{GROUPS} M. Moussa\"{\i}d, N. Perozo, S. Garnier, D. Helbing, and G. Theraulaz (2010) The walking behaviour of pedestrian social groups and its impact on crowd dynamics. {\it PLoS One} {\bf 5}(4), e10047.


\bibitem{Fences} For a photograph of the triangular fences see \url{http://loveparade2010doku.files.wordpress.com/2010/07/googlemaps_bauzaeune_rampe.jpg}.

\bibitem{foodstand} Foodstand on the ramp, see \url{http://www.youtube.com/watch?v=BhFoNb_lsO4} and \url{http://www.youtube.com/watch?v=dP-5VOCU30U}

\bibitem{policecars} Obstacles on the ramp and change of police shifts, see \url{http://live.loveparade.com/fkxt76kdrf887t/videos/chronologie/hires/06_01.mp4},\\
\url{http://www.youtube.com/watch?v=AN_8zwycDY0}

\bibitem{flowmodel} Visitor forecast of Lopavent from July 8, 2010, see\\
\url{http://loveparade2010doku.files.wordpress.com/2011/08/bewegungsmodell.jpg},\\
\url{http://loveparade2010doku.files.wordpress.com/2011/08/bewegungsmodell.pdf}, and Ref. \cite{Leaks}.

\bibitem{Schaller} Interview with the organizer of the Love Parade, see \url{http://www.youtube.com/watch?v=wsIvv6GQVkg}

\bibitem{Bewegungsmodell} Estimated inflows and outflows based on videos of surveillance camera 13 are provided at \url{http://loveparade2010doku.wordpress.com/2011/08/06/bewegungsmodell/}

\bibitem{AccesS} The operation of the access points (isolating devices) can been seen in this video: \url{http://www.youtube.com/watch?v=DTbJ_vbT8Cw} The video also mentions communication problems and that the emergency units were not prepared for such a disaster. 

\bibitem{aerial} Aerial photographs of the Love Parade in Duisburg, see \url{http://www.aerophoto.de/album.php?id=20100724%2018%201%20Loveparade_2010&language=0}

\bibitem{Hoist} Video of the crowd on the festival area, see \url{http://www.youtube.com/watch?v=bd81KIpuVc8} and the recordings of surveillance camera 4 \cite{Surveillance}.

\bibitem{StillReport} G. K. Still, Duisburg - 24th July 2010, Love Parade Incident, see \url{http://www.derwesten-recherche.org/wp-content/uploads/2012/02/Still-Gutachten.pdf}

\bibitem{Fruin2} Fruin, J. J.,
Designing for pedestrians: A level-of-service concept,
\textit{Highway Research Record} {\bf 355} (1971) 1--15. 

\bibitem{TRIANGLE}  Surveillance videos of the ramp can be seen at
\url{http://live.loveparade.com/fkxt76kdrf887t/videos/kameras/kamera13/hires/Kamera13_1400_1420.mp4}
\url{http://live.loveparade.com/fkxt76kdrf887t/videos/kameras/kamera13/hires/Kamera13_1420_1440.mp4},
\url{http://live.loveparade.com/fkxt76kdrf887t/videos/kameras/kamera13/hires/Kamera13_1440_1500.mp4},
\url{http://live.loveparade.com/fkxt76kdrf887t/videos/kameras/kamera13/hires/Kamera13_1500_1520.mp4},
\url{http://live.loveparade.com/fkxt76kdrf887t/videos/kameras/kamera13/hires/Kamera13_1520_1540.mp4},
\url{http://live.loveparade.com/fkxt76kdrf887t/videos/kameras/kamera13/hires/Kamera13_1540_1600.mp4},
\url{http://live.loveparade.com/fkxt76kdrf887t/videos/kameras/kamera13/hires/Kamera13_1600_1620.mp4},
\url{http://live.loveparade.com/fkxt76kdrf887t/videos/kameras/kamera13/hires/Kamera13_1620_1640.mp4}.

\bibitem{RMP} D. Helbing (2001) Traffic and related self-driven many-particle systems. {\it Reviews of Modern Physics} {\bf 73}, 1067-1141. 

\bibitem{PoliceShifts} Change in police shift, see \url{http://www.youtube.com/watch?v=YGPrzaxsD7I&list=UUlmsa1MvDRyVXsvCld8LMzQ&index=2&feature=plcp} 

\bibitem{CordonWest} Police cordon in the Western tunnel, see \url{http://www.youtube.com/watch?v=B-YX7tvcVYw}, \url{http://www.youtube.com/watch?v=e_exvp1NMjw}, and
\url{http://www.youtube.com/watch?v=jXcTJfmS7RQ}.

\bibitem{Cordon3} Formation of police cordon 3 in the middle of the ramp, see \url{http://www.youtube.com/watch?v=W5BOam3eGxA} and \url{http://www.youtube.com/watch?v=joBsBYcGVsw}

\bibitem{QUEUES} Queues are forming on both sides of the narrowing created by triangular fences on the ramp, when police forces start to control in- and outflows, thereby creating a bottleneck, see
\url{http://live.loveparade.com/fkxt76kdrf887t/videos/kameras/kamera13/hires/Kamera13_1600_1620.mp4},
\url{http://www.youtube.com/watch?v=W5BOam3eGxA}, 
and \url{http://www.youtube.com/watch?v=joBsBYcGVsw}.

\bibitem{secondcordon} The second police cordon in the Eastern tunnel opens up, see \url{http://www.youtube.com/watch?v=SGx1hsOmm_M} and \url{http://live.loveparade.com/fkxt76kdrf887t/videos/kameras/kamera16/hires/Kamera16_1600_1620.mp4}

\bibitem{firststaircase} First visitors are entering the festival area via the narrow staircase, see \url{http://www.youtube.com/watch?v=KaDoWMAZYyo} and \url{http://www.youtube.com/watch?v=-CEnjxQBmf4}

\bibitem{block} Security people prevent a flow of people on the staircase, see \url{http://www.youtube.com/watch?v=lFxyl9OaqHk}

\bibitem{firstdissolves} The first police cordon in the Western tunnel opens up, see 
\url{http://live.loveparade.com/fkxt76kdrf887t/videos/kameras/kamera15/hires/Kamera15_1620_1640.mp4}

\bibitem{firstpole} First people are climbing the pole, see \url{http://www.youtube.com/watch?v=VDOlXcobbJM}, \url{http://www.youtube.com/watch?v=MvlzywaFnmc}, \url{http://www.youtube.com/watch?v=6SXgp3VlM88}

\bibitem{thirddissolves} The third police cordon dissolves, see \url{http://www.youtube.com/watch?v=_PQqBePT6ig} and
\url{http://live.loveparade.com/fkxt76kdrf887t/videos/kameras/kamera13/hires/Kamera13_1620_1640.mp4}

\bibitem{stairuse} Visitors are using the small staircase to get up to the festival areas, see \url{http://www.youtube.com/watch?v=so6-7Ezeo3U}

\bibitem{signclimb} Someone has climbed a traffic sign, see \url{http://www.youtube.com/watch?v=so6-7Ezeo3U#t=2m41s}

\bibitem{forthform} A fourth police cordon is formed in the upper area of the ramp, see 
\url{http://live.loveparade.com/fkxt76kdrf887t/videos/kameras/6ersplit/hires/6erSplit_HiRes_1620_1640.mp4}

\bibitem{mauer} Crowd shouting ``Die Mauer muss weg!'' [``We must get rid of the wall (cordon)!''], see \url{http://www.youtube.com/watch?v=Opd0rZVsspQ}

\bibitem{rampflow} Jamming on the upper part of the ramp, where visitors try to enter the festival area, see
\url{http://live.loveparade.com/fkxt76kdrf887t/videos/kameras/kamera12/hires/Kamera12_1400_1420.mp4}
\\ and \\ \url{http://live.loveparade.com/fkxt76kdrf887t/videos/kameras/kamera13/hires/Kamera13_1400_1420.mp4}

\bibitem{Boeschung} Visitors climb the slopes to reach the festival area from the ramp, see
\url{http://live.loveparade.com/fkxt76kdrf887t/videos/kameras/kamera12/hires/Kamera12_1520_1540.mp4}

\bibitem{overcomingfences} According to the YouTube description, a policeman says: ``The venue is full.'', and people overcome fences to get to the festival area from the tunnel, see
\url{http://www.youtube.com/watch?v=kOjjW7Jp_Uw} and \url{http://www.youtube.com/watch?v=jm-ScKTV6nw}

\bibitem{interrupted} Flow on staircase, see \url{http://www.youtube.com/watch?v=VDOlXcobbJM#t=4m50s}

\bibitem{scream} People scream for their lives, see \url{http://www.youtube.com/watch?v=y_agoPlP_dA#t=4m22s}

\bibitem{signbent} A bent traffic sign can be seen here: \url{http://www.youtube.com/watch?v=VDOlXcobbJM}; later on, it disappeared below the crowd, see \url{http://www.youtube.com/watch?v=rzz5geLWPV4}

\bibitem{moveon} People shout that others should move on more quickly, see \url{http://www.youtube.com/watch?v=t3nDQti-zDY} and
\url{http://www.youtube.com/watch?v=t3nDQti-zDY}.

\bibitem{lightpole} Signs of crowd turbulence around the pole, see \url{http://www.youtube.com/watch?v=h3ik6n2BPa8}

\bibitem{limiting}
\url{http://www.youtube.com/watch?v=UWXXDEZ4oKg}\\
\url{http://www.youtube.com/watch?v=YmQR6kgwSxA}\\
\url{http://www.youtube.com/watch?v=r1toPUusRGU} 

\bibitem{unconscious} An unconscious woman is carried over the crowd towards the staircase, see \url{http://www.youtube.com/watch?v=KbvDLmQTED8}

\bibitem{tunnelcrowd} Slowly moving crowd in the tunnel, see \url{http://www.youtube.com/watch?v=CPcH4zZtY7w}

\bibitem{announcement} Loud speaker announcement stating ``The venue is full.'', see \url{http://www.youtube.com/watch?v=Y85nUacO2GU}

\bibitem{coolwojtek2} Problems between staircase and tunnel, see \url{http://www.youtube.com/watch?v=8aTXT3ht8VU}

\bibitem{coolwojtek3} Many people raising their hands, hoping for help (visible in full screen mode), see \url{http://www.youtube.com/watch?v=1lHLIn78650}

\bibitem{coolwojtek4} An emergency vehicle enters the ramp, see \url{http://www.youtube.com/watch?v=XKo3rb5R_Dc}

\bibitem{italian} http://www.youtube.com/watch?v=0VEbvBMrAG8 

\bibitem{coolwojtek5} A few people crawl on top of others, see \url{http://www.youtube.com/watch?v=xxd_KlaCiNY} 

\bibitem{crawl} Videos showing people crawling or walking on others, see \url{http://www.youtube.com/watch?v=JPIn5DPInB4},
\url{http://www.youtube.com/watch?v=y_agoPlP_dA}, and \url{http://www.youtube.com/watch?v=OfQjXi3J3ns}


\bibitem{ladder} People climb up from the container in the South, see http://www.youtube.com/watch?v=2B5o2wgdHcw

\bibitem{breezer1} Loveparade 2010 Disaster FullHD.mp4 16:54 - 17:03, see \url{http://www.youtube.com/watch?v=OfQjXi3J3ns}

\bibitem{cont} People are pulled up from the container, see \url{http://www.youtube.com/watch?v=IoxPIvrFCNg} 

\bibitem{screaming} A woman screams for her life, see \url{http://www.youtube.com/verify_age?next_url=/watch%3Fv%3D3x00DBo4gb8} 

\bibitem{yell} Many people yell for help, see \url{http://www.youtube.com/watch?v=YxIZwvvhpp8}

\bibitem{kaydee4} \url{http://www.youtube.com/watch?v=t3_3UIZS3dw}


\bibitem{kaydee5} \url{http://www.youtube.com/watch?v=Kc8wEMiOxoo}

\bibitem{artofhell2} \url{http://www.youtube.com/watch?v=BbCrmUZeJoY} 

\bibitem{artofhell3} \url{http://www.youtube.com/watch?v=NLC3vyp0b9U}

\bibitem{uwedix} \url{http://www.youtube.com/watch?v=OgJYkNDiDCY} 

\bibitem{kaydee6} \url{http://www.youtube.com/watch?v=gN4NNmxtRU4}

\bibitem{mbreezer8} \url{http://www.youtube.com/verify_age?next_url=/watch%3Fv%3Ds2ywRwxfHbo} 

\bibitem{artofhell4} \url{http://www.youtube.com/watch?v=QcAVAYDolKc}

\bibitem{kaydee7} \url{http://www.youtube.com/watch?v=JfaLr_Y4U18}

\bibitem{artofhell5} \url{http://www.youtube.com/watch?v=-iCA6g984aY} 

\bibitem{artofhell6} \url{http://www.youtube.com/watch?v=LXLuoqNDJ-A} 

\bibitem{kaydee8} \url{http://www.youtube.com/watch?v=FYXQLgd_VA8}

\bibitem{rkjorge1} \url{http://www.youtube.com/watch?v=Nb0m_n0KGms} 

\bibitem{artofhellmore} \url{http://www.youtube.com/watch?v=-7WDbM4EonY} \\
\url{http://www.youtube.com/watch?v=zCRGtjcbyw8}\\
\url{http://www.youtube.com/watch?v=gEa9_k-Th7U}\\
\url{http://www.youtube.com/watch?v=_CiEkTPk9BQ}

\bibitem{success} Operation room of the city of Duisburg considers the Love Parade as success as late as 17:15, see 
\url{http://www.youtube.com/watch?v=S9ILNAv0J1A} and \url{http://www.youtube.com/watch?v=FHD8aqsCr9U} 

\bibitem{rkjorge2} \url{http://www.youtube.com/watch?v=qQbto4MNqOU}

\bibitem{rkjorge3} \url{http://www.youtube.com/watch?v=WR56iEfBQoo} 

\bibitem{weiter} Videos of the continuation of the Love Parade after the Crowd Disaster, see \url{http://www.youtube.com/watch?v=Dz8dID-xTBo}

\bibitem{RTLIIPart1} RTL documentation, see \url{http://www.youtube.com/watch?v=fy1NDX_nA3M}

\bibitem{stairfalling}
Falling from staircase as assumed cause of the crowd disaster, see:
\url{http://www.n-tv.de/panorama/Loveparade-endet-im-Unglueck-18-Menschen-sterben-article1127116.html},
\url{http://www.youtube.com/watch?v=IreaH16lm_c},
\url{http://www.youtube.com/watch?v=Dz8dID-xTBo} 

\bibitem{PanicCause} Mass panic (stampede) as assumed cause of the crowd disaster, see:\\
\url{http://www.focus.de/panorama/welt/loveparade-mindestens-15-tote-nach-massenpanik_aid_533916.html}\\
\url{http://www.welt.de/die-welt/politik/article8627956/15-Tote-bei-der-Love-Parade.html}\\
\url{http://www.videoportal.sf.tv/video?id=b5b05584-57b7-42fe-bbb3-5c28d96ff6aa}

\bibitem{WrongBehavior} Wrong behavior of people assumed as reason for the crowd disaster:
\url{http://mp3-download.swr.de/swr1/bw/leute/michael-schreckenberg-stauexperte-und-panikforscher.6444m.mp3} and
\url{http://www.youtube.com/watch?v=YCxAfXjXom4} 

\bibitem{asphyxia} Wikipedia article on stampedes, see \url{http://en.wikipedia.org/wiki/Stampede}

\bibitem{MadnessOfCrowds} C. MacKay, {\it Extraordinary Popular Delusions and The Madness of Crowds}  (Harriman House, Petersfield, Hampshire, 2003).

\bibitem{Bagdad} For information on the stampede in Baghdad on August 25. 2005, see \url{http://en.wikipedia.org/wiki/2005_Baghdad_bridge_stampede}

\bibitem{Chicago} For information on the Chicago night club disaster on February 17, 2003, see \url{http://en.wikipedia.org/wiki/2003_E2_nightclub_stampede}s

\bibitem{nopanic1} J. P. Keating, The myth of panic, {\it Fire Journal}, 57-61+147 (May/1982).

\bibitem{nopanic2} I. Helsloot and A. Ruitenberg, Citizen response to disasters: a survey of literature and some practical implications, {\it Journal of Contingencies and Crisis Management} {\bf 12}(3), 98-111 (2004).

\bibitem{EarlyProblems} Early overcrowding outside the festival area, see \url{http://www.youtube.com/watch?v=BRUlHnvJl-Q}, \url{http://www.youtube.com/watch?v=Opd0rZVsspQ}, \url{http://www.youtube.com/watch?v=2aJojVY0E9A}, and  \url{http://www.youtube.com/watch?v=DTbJ_vbT8Cw}. See also Ref. \cite{floats}.

\bibitem{queuing} D. Helbing, Models for pedestrian behavior.
Pages 93-98 in: {\it Natural Structures. Principles, Strategies, and Models in Architecture and Nature}, Part II (Sonderforschungsbereich 230, Stuttgart, 1992); see also Ref. \cite{TranSci} and D. Helbing, {\it Verkehrsdynamik} (Springer, Berlin, 1997).

\bibitem{turbulence} D. Helbing, A. Johansson, and H. Z. Al-Abideen, 
The dynamics of crowd disasters: An empirical study,
\textit{Physical Review E} {\bf 75}, 046109 (2007).

\bibitem{WenjianAnders} W. Yu and A. Johansson, Modeling crowd turbulence by many-particle simulations. {\it Physical Review E} {\bf 76}, 046105 (2007).

\bibitem{inflow} Surveillance videos by the organizers of the Love Parade in Duisburg, showing the situation on the upper part of the ramp: \\
\url{http://live.loveparade.com/fkxt76kdrf887t/videos/kameras/kamera12/hires/Kamera12_1400_1420.mp4}\\
\url{http://live.loveparade.com/fkxt76kdrf887t/videos/kameras/kamera12/hires/Kamera12_1420_1440.mp4}\\
\url{http://live.loveparade.com/fkxt76kdrf887t/videos/kameras/kamera12/hires/Kamera12_1440_1500.mp4}\\
\url{http://live.loveparade.com/fkxt76kdrf887t/videos/kameras/kamera12/hires/Kamera12_1500_1520.mp4}\\
\url{http://live.loveparade.com/fkxt76kdrf887t/videos/kameras/kamera12/hires/Kamera12_1520_1540.mp4} etc.

\bibitem{Rampe} Videos showing the overcrowded ramp and visitors moving up the slopes left and/or right of the ramp towards the festival area:\\
\url{http://www.youtube.com/watch?v=qzNrGBBw-nA}\\
\url{http://www.youtube.com/watch?v=loX7-LviRr0}\\
\url{http://www.youtube.com/watch?v=AN_8zwycDY0}\\
\url{http://www.youtube.com/watch?v=fjQMaO7IBmQ},
and \url{http://www.youtube.com/watch?v=36YlTJTrRAc}

\bibitem{floats} 	
Accordingly to our notes, early crowd turbulence on the upper part of the ramp leading onto the festival area was visible on the video \url{http://video.web.de/watch/7678046}, when we accessed it around May 9, 2011, but it does not seem to be available in the Web anymore. According to the archive provided by the Wayback Machine at \url{http://www.archive.org/}, this video was apparently linked to an article entitled ``NRW erhebt schwere Vorw\"urfe gegen Veranstalter von Love Parade'', which was published by web.de. 

\bibitem{FLoaT} A float is slowly passing the ramp when people start using the slopes to get to the festival area, see \url{http://live.loveparade.com/fkxt76kdrf887t/videos/chronologie/hires/10.mp4}. The interaction between the crowd and the floats can be seen here: \url{http://live.loveparade.com/fkxt76kdrf887t/videos/kameras/kamera12/hires/Kamera12_1520_1540.mp4}

\bibitem{containerclimb} First person climbs the container around 16:24, see \url{http://www.youtube.com/watch?v=_PQqBePT6ig} and
\url{http://www.youtube.com/watch?v=yjefyv8ShfY}.

\bibitem{helping} People offering help (e.g. water), see \url{http://www.youtube.com/watch?v=VDOlXcobbJM}
and \url{http://www.youtube.com/watch?v=C0zLKDEglOs}

\bibitem{ADDitional} Group pushing to the tunnel, see \url{http://www.youtube.com/watch?v=0VEbvBMrAG8} around time 1:07. 

\bibitem{BlastAway} ``The staircase should have been blasted away'', see \url{http://www.sueddeutsche.de/wissen/loveparade-experte-zur-ungluecksursache-die-treppe-haette-man-sprengen-muessen-1.979428-2}; \url{http://www.tagesanzeiger.ch/panorama/vermischtes/Die-Treppe-haette-man-sprengen-muessen/story/10003282}

\bibitem{FOTO} For photographs of the accident area see \url{http://loveparade2010doku.files.wordpress.com/2011/08/ungluecksstelle-loveparade6.jpg}, 
\url{http://hvg.hu/nagyitas/20100726_love_parade_duisburg_nagyitas#kep8},
\url{http://farm5.static.flickr.com/4134/4912001905_3871745170_o.jpg}. For further photographs of the Love Parade disaster, see 
\url{http://loveparade2010doku.wordpress.com/bilder/} and \url{http://www.wz-newsline.de/home/panorama/die-tragoedie-auf-der-loveparade-2010-1.21599}

\bibitem{FallingFromStairs} Someone falling down, when trying to climb the staircase from the side, see \url{http://www.youtube.com/watch?v=JigIrMkWdYY},
\url{http://www.youtube.com/user/rkjorge70}, and \url{http://www.youtube.com/user/LoveparadeDuisburg?feature=watch}

\bibitem{WirNehmenDieTreppe} ``Let us take the staircase'' (``Wir nehmen die Treppe''), see \url{http://loveparade2010doku.wordpress.com/2010/07/28/loveparade-2010-zeitablauf-sperrungen-und-durchlassungen/}

\bibitem{ClimbFlow} The flow of people on the staircase is considerably obstructed by people, who climb the staircase from the side, see \url{http://www.youtube.com/watch?v=G-mfsxz2k2g}

\bibitem{stumbling} Falling due to the turbulent waves in the crowd, see \url{http://www.youtube.com/watch?v=S9ILNAv0J1A} 

\bibitem{Domino} Domino effect in the crowd and involuntary stepping on others to survive, see \url{http://www.wdr.de/mediathek/html/regional/2011/07/19/lokalzeit-duisburg-loveparade.xml}

\bibitem{turbulenceInNews} Crowd turbulence as assumed cause of the crowd disaster, see \url{http://www.pressetext.com/news/20100729019}

\bibitem{eyewitness} Eyewitnesses describing the cause of the disaster, see \url{http://www.youtube.com/watch?v=b79riRVinJs} and \url{http://www.youtube.com/watch?v=S9ILNAv0J1A}

\bibitem{TURbu} For signs of crowd turbulence on lower part of the ramp see \url{http://live.loveparade.com/fkxt76kdrf887t/videos/kameras/kamera13/hires/Kamera13_1620_1640.mp4}, \url{http://www.youtube.com/watch?v=JigIrMkWdYY}, and  \url{http://www.youtube.com/watch?v=-2onoJLq2-8}.
It seems, however,  that crowd turbulence occurred even earlier in the upper part of the ramp; a video showing this with the file name \url{http://video.web.de/watch/7678046} was apparently linked to the following web.de article:
NRW erhebt schwere Vorwürfe gegen Veranstalter von Love Parade Video, but it is not accessible anymore.

\bibitem{LyingOnGround} Videos of the area where the victims were located, see 
\url{http://www.youtube.com/watch?v=WR56iEfBQoo}, \url{http://www.youtube.com/watch?v=qbQr5yAjqxs},
and \url{http://www.youtube.com/watch?v=FYXQLgd_VA8},
\url{http://www.youtube.com/watch?v=JPIn5DPInB4}, and Refs. \cite{hell} and \cite{rkjorge70}.

\bibitem{wonder} \url{http://loveparade2010doku.wordpress.com/2010/07/28/loveparade-2010-zeitablauf-sperrungen-und-durchlassungen/} (nach 16:57, "Einschub")

\bibitem{manhole} Pictures of the broken manhole, see \url{http://www.dailymotion.com/video/xfl3rc_rtl2-100tage-loveparade-ausschnitte_news} (cut version of \cite{RTL}).

\bibitem{noturb} Video recordings of the situation around the emergency vehicle on the ramp, see 
\url{http://www.youtube.com/watch?v=0VEbvBMrAG8},
\url{http://www.youtube.com/watch?v=yNuh9Hk_tgM},
\url{http://www.youtube.com/watch?v=Z5mf7imNrIY},
\url{http://www.youtube.com/watch?v=G-mfsxz2k2g},
\url{http://www.youtube.com/watch?v=dP-5VOCU30U}, and
\url{http://www.youtube.com/watch?v=rzz5geLWPV4}

\bibitem{comparison} For some aerial photographs of the festival area of the Love Parade in Berlin see \url{http://www.skyscrapercity.com/showpost.php?s=e6cd40536463e09d85e914ba4372578f&p=60946437&postcount=44}

\bibitem{PrivComm} Private communication; for further eye witness reports see \url{http://loveparade2010doku.wordpress.com/2010/07/27/erinnerungen-von-augenzeugen-zusammengefasst-und-verlinkt/}

\bibitem{counterflows} D. Helbing, I. Farkas, and T. Vicsek (2000) Freezing by heating in a driven mesoscopic system. Physical Review Letters 84, 1240-1243.

\bibitem{Augenzeuge} Eye witness report, Anlage 68 [attachment no. 68] at \url{https://www.duisburg.de/ratsinformationssystem/bi/vo0050.php?__kvonr=20056110&voselect=20049862}

\bibitem{helicopter} Surveillance by police and helicopter, see \url{http://www.youtube.com/watch?v=t3_3UIZS3dw},  \url{http://www.youtube.com/watch?v=8aQuTaMbS38}, 
\url{http://www.youtube.com/watch?v=YmQR6kgwSxA} and Ref. \cite{coolwojtek5}.

\bibitem{late}
Situation after 17:20 on the main ramp, entering of a loudspeaker vehicle, and activities in the tunnel, see 
\url{http://www.youtube.com/watch?v=XgCByAtNXmY},
\url{http://www.youtube.com/watch?v=SqqdnzAsF60} (removed),
\url{http://www.youtube.com/watch?v=ffqkFOXX3AM}, 
\url{http://www.youtube.com/watch?v=y4W3xb-SMDs}, and
\url{http://www.youtube.com/watch?v=j11330qbfb4}

\bibitem{firstattempts} The police attempts to direct the crowd on the ramp to the upper end, see \url{http://www.youtube.com/watch?v=3XFTVWUN8nw}


\bibitem{Physica_A} D. Helbing and C. K\"uhnert, Assessing interaction networks with applications to catastrophe dynamics and disaster management. {\it Physica A} {\bf 328}, 584-606 (2003).

\bibitem{NoSenseOfEmergency} The seriousness of the situation is not recognized, despite the extremely crowded situation around the pole, the container, and the staircase, see \url{http://www.youtube.com/watch?v=9E32qAeOB0k},  \url{http://www.youtube.com/watch?v=-2onoJLq2-8}, \url{http://www.youtube.com/watch?v=fL_1reQrERg}, \url{http://www.youtube.com/watch?v=N795HhESDkw} and \url{http://www.youtube.com/watch?v=OfQjXi3J3ns}; also note that most videos do not record the accident area for a long time---the filmers do not seem to notice the location where the conditions was most critical.

\bibitem{FollowPolice} Many visitors follow the leaving emergency vehicle up the ramp, see \url{http://www.youtube.com/watch?v=rzz5geLWPV4}
and \url{http://www.youtube.com/watch?v=dP-5VOCU30U} 

\bibitem{Triage} O. Ackermann {\it et al.}, Patient care at the 2010 Love Parade in Duisburg, Germany:
Clinical experiences, {\it Deutsches \"Arzteblatt International} {\bf 108}(28/29), 483-489 (2011), see \url{http://www.ncbi.nlm.nih.gov/pmc/articles/PMC3149288/}; see also \url{http://www.youtube.com/watch?v=S9ILNAv0J1A}.
 
\bibitem{trafficcircle} Y. Sugiyama {\it et al.} Traffic jams without bottlenecks---experimental evidence for the physical mechanism of the formation of a jam. {\it New Journal of Physics} {\bf 10}, 033001.

\bibitem{Hardin} G. Hardin, The tragedy of the commons. {\it Science} {\bf 162}, 1243-1248 (1968).

\bibitem{revolutions1} D. Helbing and A. Johansson, Cooperation, norms, and revolutions: A unified game-theoretical approach. {\it PLoS ONE} {\bf 5}(10), e12530 (2010).
\bibitem{revolutions2} D. Helbing and A. Johansson, Evolutionary dynamics of populations with conflicting interactions: Classification and analytical treatment considering asymmetry and power. {\it Physical Review E} {\bf 81}, 016112 (2010).
\bibitem{revolutions3} D. Helbing, FuturICT---New science and technology to manage our complex, strongly connected world, preprint \url{http://arxiv.org/abs/1108.6131}.

\bibitem{IRGC} D. Helbing (2010) Systemic Risks in Society and Economics. International Risk Governance Council (irgc), see \url{http://www.irgc.org/IMG/pdf/Systemic_Risks_Helbing2.pdf}

\bibitem{queen} In a letter dated July 22, 2009, from the (British) Royal Academy to Her Majesty, the Queen, the following explanation was given for the financial crisis, see  \url{http://www.euroresidentes.com/empresa_empresas/carta-reina.pdf}: 
``... where was the problem? Everyone seemed to be doing their own job properly on its own merit. And according to standard measures of success, they were often doing it well. The failure was to see how collectively this added up to a series of interconnected imbalances over which no single authority had jurisdiction. ...  Individual risks may rightly have been viewed as small, but the risk to the system as a whole was vast.'' This quote illustrates the mechanism of systemic instability quite well (which may be defined as a situation in which things can get totally out of control, even when everyone is trying to do his or her job in an adequate way).

\bibitem{Spread1}  D. Helbing, H. Ammoser, and C. K\"uhnert, Disasters as extreme events and the importance of network interactions for disaster response management. Pages 319-348 in S. Albeverio, V. Jentsch, and H. Kantz (eds.) {\it Extreme Events in Nature and Society} (Springer, Berlin, 2005).

\bibitem{Spread2}
   K. Peters, L. Buzna, and D. Helbing, Modelling of cascading effects and efficient response to disaster spreading in complex networks. {\it Int. J. Critical Infrastructures} {\bf 4}(1/2), 46-62 (2008).

\bibitem{BlackoutCascades}     I. Simonsen, L. Buzna, K. Peters, S. Bornholdt and D. Helbing (2008) Transient dynamics increasing network vulnerability to cascading failures. {\it Physical Review Letters} {\bf 100}, 218701. 

\bibitem{metropole} \url{http://www.metropoleruhr.de/en/home/ruhr-metropolis/data-facts.html}

\bibitem{LateStartNoSeparation} Jamming due to the later start of the Love Parade and criticism of the lack of flow separation, see \url{http://www.wdr.de/mediathek/html/regional/2010/07/26/lokalzeit-duisburg-loveparade-floatveranstalter.xml} 

\bibitem{NotExpected} The emergency forces were not prepared for the large number of injuries, see \url{http://www.youtube.com/watch?v=DTbJ_vbT8Cw} 

\bibitem{VisualAnalytics} A. Johansson, {\it Data-Driven Modeling of Pedestrian Crowds} (VDM
Publishers Dr. Mueller, 2009), see \url{http://tud.qucosa.de/fileadmin/data/qucosa/documents/2090/dissertation_johansson_tudresden.pdf} and also \url{http://www.crowdvision.co.uk}

\bibitem{QuickDisasterResponse}
   L. Buzna, K. Peters, H. Ammoser, C. K\"uhnert, and D. Helbing (2007) Efficient response to cascading disaster spreading. Physical Review E 75, 056107.

\bibitem{moveup} Police supports people moving up the narrow staircase from 16:35, see \url{http://www.youtube.com/watch?v=AYjlXhPcF4s}

\bibitem{circular} Comparison of the flow organization of the Love Parade (characterized by counterflows) with an alternative, circular flow concept, see 
\url{http://loveparade2010doku.wordpress.com/2010/07/28/das-umgesetzte-wegekonzept-der-loveparade-2010-und-eine-von-vielen-moglichen-alternativen-im-vergleich/}

\bibitem{trust} Mitarbeiter der Stadt Duisburg werden angefeindet [Staff of the City of Duisburg publicly attacked], see \url{http://www.derwesten.de/staedte/duisburg/mitarbeiter-der-stadt-duisburg-werden-angefeindet-id3319267.html}.

\bibitem{Hilbert} D. Helbing, Accelerating scientific discovery by formulating grand scientific challenges, to appear in {\it EPJ Special Topics} (2012).

\end{thebibliography}

\end{document}